\documentclass[aps,english, pre, preprint,longbibliography,onecolumn,superscriptaddress]{revtex4-1}
\usepackage{bm}
\usepackage{mathrsfs}
\usepackage{latexsym,amsmath}
\usepackage{amsbsy}
\usepackage{amssymb}
\usepackage{hyperref}
\usepackage[pdftex]{graphicx}
\usepackage[usenames]{color}
\begin{document}
\title{Mapping high-growth phenotypes in the flux space of microbial metabolism}

%oriol
\author{Oriol G{\"u}ell}
\affiliation{Departament de Qu{\'\i}mica F{\'\i}sica, Universitat de Barcelona, Mart\'{\i} i Franqu\`es 1, 08028 Barcelona, Spain}

%massucci
\author{Francesco Alessandro Massucci}
\affiliation{Departament d'Enginyeria Qu\'{\i}mica, Universitat Rovira i Virgili, 43007 Tarragona, Spain}

%fontclos
\author{Francesc Font-Clos}
\affiliation{Centre de Recerca Matem\`atica, Edifici C, Campus Bellaterra, E-08193 Bellaterra, Barcelona, Spain}
\affiliation{Departament de Matem\`atiques, Universitat Aut\`onoma de Barcelona, Edifici C, Campus Bellaterra, E-08193 Bellaterra, Barcelona, Spain}

%sagues
\author{Francesc Sagu{\'e}s}
\affiliation{Departament de Qu{\'\i}mica F{\'\i}sica, Universitat de Barcelona, Mart\'{\i} i Franqu\`es 1, 08028 Barcelona, Spain}

%mangeles
\author{M. \'Angeles Serrano}
\email{marian.serrano@ub.edu\\All authors contributed equally to this work}

\affiliation{Departament de F{\'\i}sica Fonamental, Universitat de Barcelona, Mart\'{\i} i Franqu\`es 1, 08028 Barcelona, Spain}

%\linenumbers
%% ABSTRACT %%
\begin{abstract}
Experimental and empirical observations on cell metabolism cannot be understood as a whole without their integration into a consistent systematic framework. However, the characterization of metabolic flux phenotypes is typically reduced to the study of a single optimal state, like maximum biomass yield that is by far the most common assumption. Here we confront optimal growth solutions to the whole set of feasible flux phenotypes (FFP), which provides a benchmark to assess the likelihood of optimal and high-growth states and their agreement with experimental results. In addition, FFP maps are able to uncover metabolic behaviors, such as aerobic fermentation accompanying exponential growth on sugars at nutrient excess conditions, that are unreachable using standard models based on optimality principles. The information content of the full FFP space provides us with a map to explore and evaluate metabolic behavior and capabilities, and so it opens new avenues for biotechnological and biomedical applications.
\end{abstract}

\maketitle

\section{Introduction}
Growing evidence suggests that metabolism is a dynamically regulated system that reorganizes under evolutionary pressure to safeguard survival~\cite{Ibarra:2002,blank2005}. This adaptability implies that metabolic phenotypes directly respond to environmental conditions. For instance, unicellular organisms can be stimulated to proliferate by controlling the abundance of nutrients available. In rich media, cells reproduce as quickly as possible by fermenting glucose, a process which produces high specific growth rates as well as large quantities of excess carbon in the form of ethanol and organic acids~\cite{Frick:2005}, a process known as the Crabtree effect~\cite{Deken:1996,Paczia:2012}. To survive the scarcity of nutrients during starvation periods, glycolysis is hypothesized to switch to oxidative metabolism, which no longer maximizes the specific growth rate, but instead the ATP yield needed for cellular processes. Cells of multicellular organisms show similar metabolic phenotypes, relying primarily on oxidative phosphorylation when not stimulated to proliferate and changing to non-oxidative glycolytic metabolism during cell proliferation, even if this process --known in cancer cells as the Warburg effect~\cite{VanderHeiden:2009}-- is much less efficient at the level of energy yield.

These metabolic phenotypes are captured by computational approaches like Flux Balance Analysis~\cite{Orth:2010} (FBA) that has been applied to high-quality genome-scale metabolic network reconstructions~\cite{overbeek05, Blattner1997, Feist:2007, oh07, mo09} to estimate the fluxes of biochemical reactions at steady state. Compliant with stoichiometric mass balance constraints and with imposed upper and lower bounds for nutrients, FBA determines the flux distribution that optimizes a biological objective such as specific growth rate, biomass yield, ATP yield or the rate of production of a biotechnologically important metabolite. This important tool has been used to predict the growth rate of organisms and to analyze their viability~\cite{Almaas:2005,Suthers:2009}. Minimization of metabolic adjustment (MOMA)~\cite{Segre:2002}, which identifies a single suboptimal point in flux space, has been proposed as an alternative option for perturbed metabolic networks not exposed to long-term evolutionary pressure. In any case, the identified solutions are frequently inconsistent with the biological reality since no single objective function describes successfully the variability of flux states under all environmental conditions~\cite{Schuetz:2007,Schuetz:2012}, and in fact the highest accuracy of FBA predictions is achieved whenever the most relevant objective function is tailored to particular environmental conditions according to the empirical evidence for a very specific metabolic phenotype. For instance, FBA maximization of growth rate, by far one of the most common assumption, requires either a rich medium or a manual limitation of the oxygen uptake to a physiological enzymatic limit to mimic the observed fermentation of glucose to formate, acetate, or ethanol typical of proliferative metabolism, while in minimal medium optimization of growth rate relies primarily on oxidative phosphorylation, which increases ATP production converting glucose to carbon dioxide, as in starvation metabolism.

Along optimal metabolic phenotypes, there is however a whole space of possible states non-reachable by invoking optimality principles that prevent non-optimal or typical biological states. Optimization of a biological function in the absence of {\it a priori} biological justification, like what happens under conditions for proliferative or starvation metabolism, may indeed weaken {\it in silico} predictions. Elementary flux modes~\cite{Schuster:1994,Schuster:2000} --non-decomposable steady-state pathways through a metabolic network such that any possible pathway can be described as a non-negative linear combination of them-- provide a view on the flux space without the requirement of any optimality function. However, calculation of all elementary flux modes for an entire network is computationally very demanding due to the combinatorial explosion of their number with increasing size of the network~\cite{Acuna:2010}. For instance, the core metabolism of {\it Escherichia coli} in~\cite{orth2009} consists of around 271 million elementary flux modes~\cite{Jungreuthmayer:2012}. To overcome this handicap, recent advances avoid the comprehensive enumeration of elementary flux modes using instead a sample of the available elementary flux mode solution space~\cite{Gebauer:2012}. Even admitting one is able to enumerate all elementary flux modes, it is however impossible to assess the likelihood of observing a given linear combination of them in a typical phenotype. Further, elementary flux modes cannot capture changes associated with reaction fluxes being capped for whichever physiological reason (Fig.~S6 in Supporting Information (SI)). On top, due to functional redundancy, the expansion of possible metabolic pathways in elementary flux modes is not unique. Therefore, enumeration of the elementary flux modes is not as insightful as characterizing the whole phenotypic space, albeit requiring a comparable computational complexity.

Here, we introduce an alternative approach that estimates directly the feasible flux phenotypic (FFP) space using a mathematically well characterized sampling technique which enables the analysis of feasible flux states in terms of their likelihood. We use it to confront optimal growth rate solutions with the whole set of feasible flux phenotypes of {\it Escherichia coli} ({\it E. coli}) core metabolism in minimal medium. The FFP space provides a reference map that helps us to assess the likelihood of optimal and high-growth states. We quantitatively and visually show that optimal growth flux phenotypes are eccentric with respect to the bulk of states, represented by the feasible flux phenotypic mean, which suggests that optimal phenotypes are uninformative about the more probable states, most of them low growth rate. We propose feasible flux phenotypic space eccentricity of experimental data as a standard tool to calibrate the deviation of optimal phenotypes from experimental observations. Finally, the analysis of the entire high-biomass production region of the feasible flux phenotypic space unveils metabolic behaviors observed experimentally but unreachable by models based on optimality principles, which forbid aerobic fermentation -a typical pathway utilization of proliferative metabolism- in minimal medium with unlimited oxygen uptake.

\section{Materials and Methods}
The FFP space, also termed the flux cone~\cite{Schilling:2000}, of a metabolic model in a specific environment has been explored using different sampling techniques~\cite{Price:2004, Wiback:2004, Almaas:2004}. Here, we use the Hit-And-Run (HR) algorithm to explore the FFP space, tailoring it to enhance its sampling rate and to minimize its mixing time. We refer the interested reader to~\cite{Massucci:2013}, where our implementation was first introduced, stating here only the key points and ideas.

\subsection{Hit-And-Run algorithm to sample the space of feasible metabolic flux solutions}
We start by noticing that all points in the FFP space must simultaneously satisfy mass balance conditions and uptake limits for internal and exchanged metabolites, respectively. The former requirement defines a set of homogeneous linear equalities, whose solution space is $K$, while the latter defines a set of linear inequalities, whose solutions lie in a convex compact set $V$. From a geometrical point of view, the FFP space is thus given by the intersection $S = K\cap V$. A key step of our approach consists in realizing that one can directly work in $S$ by sampling $V$ in terms of a basis spanning $K$. This allows to retrieve all FFPs that satisfy mass balance in the medium conditions under consideration, without rejection. Additionally, sampling in $S$ allows to perform a drastic dimensional reduction and to decrease considerably the computation time. Indeed, assuming to have $N$ reactions, $I$ internal metabolites, and $E$ exchanged metabolites ($N>I+E$), one has that $S \subset\mathbb{R}^{N-I}$, which is typically a space with greatly reduced dimensionality with respect to $V \subset \mathbb{R}^N$.

Once a basis for $K$ is found, the main idea behind HR is fairly simple. Given a feasible solution $\boldsymbol{\nu}_\mathrm{o} \in S$, a new, different feasible solution $\boldsymbol{\nu}_\mathrm{n} \in S$  can be obtained as follows:
\begin{enumerate}
\item {Choose a random direction $\boldsymbol{u}$ in $\mathbb{R}^I$}
\item {Draw a line $\ell$ through $\boldsymbol{\nu_o}$ along direction $\boldsymbol{u}$:
\[
\ell: \boldsymbol{\nu_o} + \lambda \boldsymbol{u},\qquad \lambda \in \mathbb{R}
\]}
\item {Compute the two intersection points  of $\ell$ with the boundary of $S$, parametrized by $\lambda=\lambda_{-},\lambda_{+}$:
\begin{alignat*}{1}
\boldsymbol{\nu_-} &= \boldsymbol{\nu_o} + (\lambda_{-}) \boldsymbol{u}
\\
\boldsymbol{\nu_+} &= \boldsymbol{\nu_o} + (\lambda_{+}) \boldsymbol{u}
\end{alignat*}}
\item {Choose a new point $\boldsymbol{\nu}_{\mathrm{n}}$ from $\ell$, uniformly at random between $\boldsymbol{\nu}_{\mathrm{-}}$ and $\boldsymbol{\nu}_{\mathrm{+}}$. In practice, this implies choosing a value $\lambda_{\mathrm{n}}$ in the range $(\lambda_\mathrm{-},\lambda_\mathrm{+})$ uniformly at random, and then 
\[
\boldsymbol{\nu}_{\mathrm{n}} \equiv \boldsymbol{\nu}_\mathrm{o} + \lambda_{\mathrm{n}} \boldsymbol{u}
\]}
\end{enumerate}
This procedure is repeated iteratively so that, given an initial condition, the algorithm can produce an arbitrary number of feasible solutions (see Fig.~S4 in SI for an illustrative representation of the algorithm). The initial condition, which must be a feasible metabolic flux state itself ({\em i.e.} it must belong to $S$), is obtained by other methods. We used and recommend MinOver, see~\cite{Massucci:2013, Krauth:1987}, but any other technique is valid. In particular, in cases where small samples of the FFP space have been already obtained by other sampling techniques, such points can be used to feed the HR algorithm and produce a new, larger sample.

It was proven that HR converges towards the uniform sampling of $S$~\cite{Lovasz1999} and we took several measures to ensure that this was the case in our implementation (Fig.~S4 in SI). For each model, we initially created samples of size $ 1.3 \cdot10^9$, giving rise to a final set of $10^6$ feasible solutions, uniformly distributed along the whole FFP space.

\subsection{Principal Component Analysis of the profile correlation matrix}
Compared to phenotypic optimisation or, {\em e.g.}, elementary flux modes, FFP sampling has the advantage of allowing the computation of reaction pairs correlations. These may be exploited to detect how global flux variability emerges in the system through principal component (PC) analysis\cite{Pearson1901, Jolliffe2002} and to quantify, in turn, the closeness of optimal phenotypes to the bulk of the FFP. In what follows we briefly describe the method, while an illustrative example is provided in Fig.~S5 in SI.

To perform such study, we start by writing down the matrix $C_{ij}$ of correlations between all reaction pairs $i, j$. In doing this, we measure how much the variability of reaction flux $\nu_i$ affects the flux $\nu_j$ (and viceversa). In mathematical terms, for each pair of reactions $i,j$, we have:
\begin{equation}
C_{ij} = \frac{\langle \nu_i \nu_j\rangle - \langle \nu_i\rangle\langle \nu_j\rangle}{\sqrt{\left(\langle \nu_i^2\rangle-\langle \nu_i\rangle^2\right)\left(\langle \nu_j^2\rangle-\langle \nu_j\rangle^2\right)}},
\end{equation}
where $\langle\ldots\rangle$ denotes an average over the sampled set and the denominator of the fraction is simply the product of the standard deviations of $\nu_i$ and $\nu_j$. We plot such matrix in Fig.\ref{fig1}e.

Matrix $\boldsymbol{C}$ is real and symmetric by definition and, thus, diagonalizable. This means that, for every eigenvector $\rho_\kappa$, one has $\boldsymbol{C}\rho_\kappa = \lambda_\kappa \rho_\kappa$.  Note that matrix $\boldsymbol{C}$ describes paired flux fluctuations, in a reference frame centered on the mean flux vector. The eigenvectors $\rho_\kappa$ of $\boldsymbol{C}$ express, in turn, the directions along which such fluctuations are taking place. In particular, the eigenvectors $\rho_1, \rho_2$ associated with the first two largest (in modulo) eigenvalues dictate the two directions in space where the sampled FFP displays the greatest variability (see Fig. S5 in SI). This implies that sampled phenotypes lie closer to the plane spanned by $\rho_1$ and $\rho_2$ than the ones produced by any other linear combination of $\boldsymbol{C}$ eigenvectors. Projecting all sampled FFP onto this plane allows thus to perform a drastic dimensional reduction yet retaining much of the original variability and allows to have a direct graphical insight on where phenotypes lie, on where the bulk of the FFP is located and on how the FBA solution compares to them. In such plot, each phenotype $\jmath$ is described by two coordinates, that may be parameterized via a radius $r_\jmath$ and an angle $\theta_\jmath$. Since the projection is normalised, it follows that $r_\jmath\leq1$. Furthermore, the closer $r_\jmath$ to one, the better the phenotype $\jmath$ is described by only looking at variability along $\rho_1, \rho_2$. As $r_\jmath$ is one at the most and since we have so many phenotypes clustered together, we chose to plot the PCAprojection by using an effective radius $r_\jmath' = -\log(r_\jmath)$ in Fig.~\ref{fig1}f. In this way we could better discriminate among different phenotypes and got a `closest to the origin, closest to the $\rho_1,\rho_2$--plane' setup.

As compared to previous works focused on characterizing the principal components of the solution space to obtain a low-dimensional decomposition of the steady flux states of the system~\cite{Barrett:2009}, our approach presents two main conceptual differences. First, the sampling method used here produces a uniform sample over the full set of feasible flux states without introducing any bias towards high-growth flux states. Second, we aim at a full description of all feasible flux states to conduct a statistical analysis of feasible phenotypes, which cannot be done by only retaining PCs. We use PC analysis to visualize the eccentricity of the FBA solution, but for all other purposes we take into account the whole set of metabolic states.

\section{Results}
\label{sec:results}
We study the full metabolic flux space of the  {\it E. coli} core metabolic model~\cite{orth2009,Orth:2010}, a condensed version of the genome-scale metabolic reconstruction {\it i}AF1260~\cite{Feist:2007} that contains $73$ central metabolism reactions and $72$ metabolites. This network is complemented with the biomass formation reaction and the ATP maintenance reaction. As in FBA, feasible flux states of a metabolic network are those that fulfill stoichiometric mass balance constraints together with imposed upper and lower bounds on the reaction fluxes. These constraints restrict the number of solutions to a compact convex set which contains all possible flux steady states in a particular environmental condition. In glucose minimal medium, the FFP space of {\it E. coli} core metabolism is determined by $70$ potentially active reactions, including biomass formation and ATP maintenance reaction, and $68$ metabolites. Note that we allow negative values for reversible reactions. We apply a fast and efficient Hit-And-Run algorithm~\cite{Massucci:2013} (see Materials and Methods) that explores the full solution space at random to produce a raw sample of $10^{9}$ feasible states from which we extract a final uniform representative set of $10^{6}$ feasible states.

Notice that our approach is suitable for genome-scale network sizes beyond the reduced size of the {\it E. coli} core model. There is not any fundamental or technical bottleneck that prevents its application to complete metabolic descriptions at the cell level since uniform samples can also be generated in genome-scale networks. We used the {\it E. coli} core metabolism due to a matter of computational time and ease of visualization.

\subsection{Optimal growth is eccentric with respect to the full FFP space}
\begin{figure}[htp]
\includegraphics[width=0.7\textwidth]{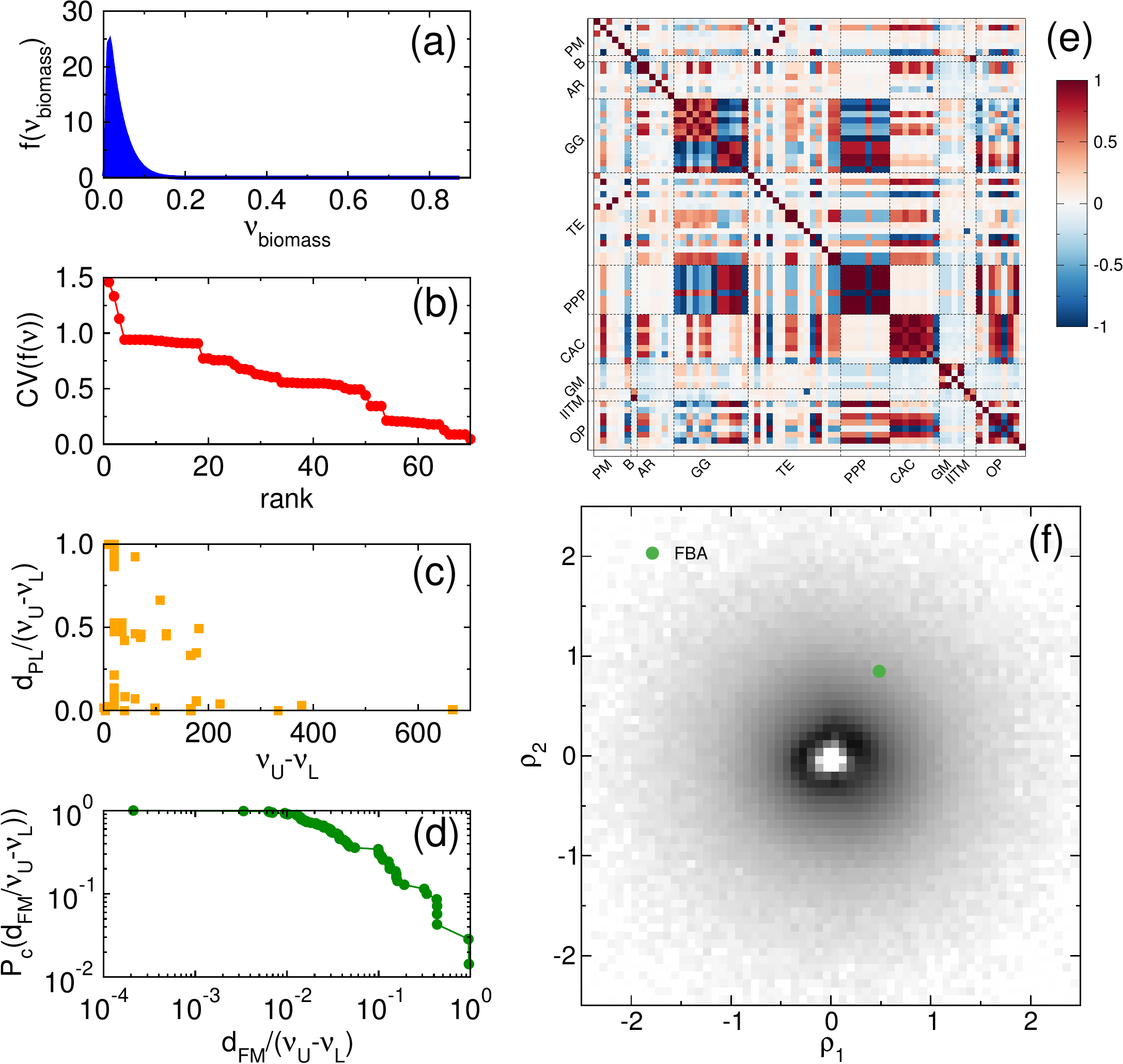}
\caption{{\bf Analysis of reaction profiles and visualization of the FFP space.} {\bf (a)} Probability density function of metabolic flux values for the biomass function in {\it E. coli} core metabolism under glucose minimal conditions. {\bf (b)} Coefficient of variation for all core reactions ranked by value. {\bf (c)} Scatterplot of distances between the more probable flux in the FFP space and the lower flux bound rescaled by flux variability range for each reaction. {\bf (d)} Complementary cumulative distribution function of distances between FBA maximal growth flux and FFP space mean flux rescaled by flux variability range for each reaction, in log-log scale. {\bf (e)} Matrix of Pearson correlation coefficients measuring the degree of linear associations between feasible fluxes of reactions (acronyms of the pathways are shown in Table S2 in SI). {\bf (f)} Projection of the FFP space onto the two principal component vectors of the correlation matrix in (e). All sampled flux phenotypes are normalized and projected along the first ($\rho_1$) and second  ($\rho_2$) principal components. The plot is in polar coordinates, with the negative logarithm of the radius. The majority of points lies in a circle close to the origin (the darker area). The FBA solution (green circle) is, conversely, rather eccentric.}
\label{fig1}
\end{figure}

From the sampled set of {\it E. coli} core metabolic states in minimal medium of glucose bounded to $10$ mmol/(gDW$\cdot$h), we collected the metabolic flux profile of each individual reaction as the set of its feasible metabolic fluxes. From such profile, we computed the probability density function $f(\nu)$ which describes the likelihood for a reaction to take on a particular flux value. As an example, see Fig.~\ref{fig1}a for the biomass function. We observe a variety of shapes (Fig.~S1 in Supporting Information (SI)), all of them low-variance, most displaying a maximum probability for a certain value of the flux inside the allowed range (notice that none of these histograms can have more than one peak due to the convexity of the steady-state flux space), and many being clearly asymmetric. 

To characterize the dispersion of the possible fluxes for each reaction, we measured its coefficient of variation $CV(f(\nu))$ calculated as the ratio between the standard deviation of possible fluxes and their average (Table S1 in SI). For all but three reversible reactions (Malate dehydrogenase, Glucose-6-Phosphate isomerase, and Glutamate dehydrogenase), the only reversible reactions having a low associated flux mean and thus a higher $CV(f(\nu))$, this metric is below one and when ranked for all reactions it steadily decreases to almost zero, Fig.~\ref{fig1}b. Interestingly, we find that this coefficient is significantly anticorrelated with the essentiality of reactions, as observed experimentally~\cite{Orth:2011} (point-biserial correlation coefficient $-0.29$ with p-value $0.01$). This means that essential reactions tend to have a highly concentrated profile of feasible fluxes. Besides, and only for the glucose transferase reaction GLCpts, we find a zero probability of having a zero flux, which is not surprising as the lower bound given by FVA is strictly greater than zero indicating that this reaction is essential for {\it E. coli} core metabolism in glucose minimal medium. The asymmetry of each profile was characterized by the distance between the more probable flux in the FFP space and the lower flux bound of the flux variability range rescaled by the flux variability range of the reaction (Table S1 in SI). In Fig.~\ref{fig1}c we show a scatterplot of values for all $68$ core reactions. Strikingly, the rescaled distances cluster in three regions around $0$, $0.5$ and $1$ forming groups of sizes $38$, $15$ and $17$ respectively. This indicates that the most probable flux is close to either the lower or upper bound or, conversely, the probability distribution function tends to be quite symmetric. Moreover, we also observe an anticorrelation between the length of the flux range and the position of the most probable flux, so that the closer is this to its maximum value the shorter is the allowed range of fluxes.  

In order to assess the likelihood of FBA maximization of the biomass reaction (FBA-MBR) (or equivalently of the growth rate) solutions in relation to typical points within the whole FFP space (typical, in our mathematical/computational context, means statistically representative in relation to the whole set of flux states contained in the FFP space), we calculated the average flux value for each reaction, that we named the mean, and compared it to the FBA optimal biomass production flux. The complementary cumulative distribution function of the distances between these two characteristic fluxes rescaled by the flux variability range of reactions is shown in Fig.~\ref{fig1}d (Table S1 in SI). We observe a broad distribution of values over several orders of magnitude with no mean value actually very close to the FBA maximal solution except for a few reactions, which typically work at maximum growth. At the other end of the spectrum, deviated reactions include for instance excretion of acetate and phosphate exchange. As a summary, we conclude that the mean and the FBA biomass optimum are rather distant, which suggests that FBA optimal states are uninformative about phenotypes in the bulk of states in the FFP space.

To visualize neatly the eccentricity of the FBA maximum growth state with respect to the bulk of metabolic flux solutions, we used Principal Component Analysis~\cite{Pearson1901, Jolliffe2002} in order to reduce the high-dimensionality of the full flux solution space projecting it onto a two-dimensional plane from the most informative viewpoint (see Materials and Methods). We took reaction profiles in pairs to calculate the matrix of Pearson correlation coefficients measuring their degree of linear association Fig.~\ref{fig1}e (Table S3 in SI). Note that an ordering of reactions by pathways allows to have a clear visual feedback of intra- and inter-pathway correlations taking place in the core metabolic network, such that clusters of highly correlated reactions appear as bigger darker squares. The two axes of our projection correspond to the two first principal components of this profile correlation matrix $\rho_1$ and $\rho_2$, which account for most of the variability in profile correlations. Each sampled metabolic flux state was rescaled as a z-score centered around the mean and projected onto these axes, as shown in the scatterplot Fig.~\ref{fig1}f in polar coordinates, where we applied a negative logarithmic transformation to the radial coordinate for ease of visualization. We see that the majority of phenotypes have a radius close to zero. Since points closer to the origin are better described by the two principal components, this implies that $\rho_1$ and $\rho_2$ capture the largest variability of the sampled FFP. Clearly, the FBA optimal growth solution is rather eccentric with respect to typical solutions, with an associated radius of $0.98$ in this representation. In fact, $97\%$ of states have a smaller radius than the optimal growth solution (see Fig.~S2 in SI). 

\subsection{The FFP space gives a benchmark to calibrate the deviation of optimal phenotypes from experimental observations}
\begin{figure}[htp]
\includegraphics[width=0.45\textwidth]{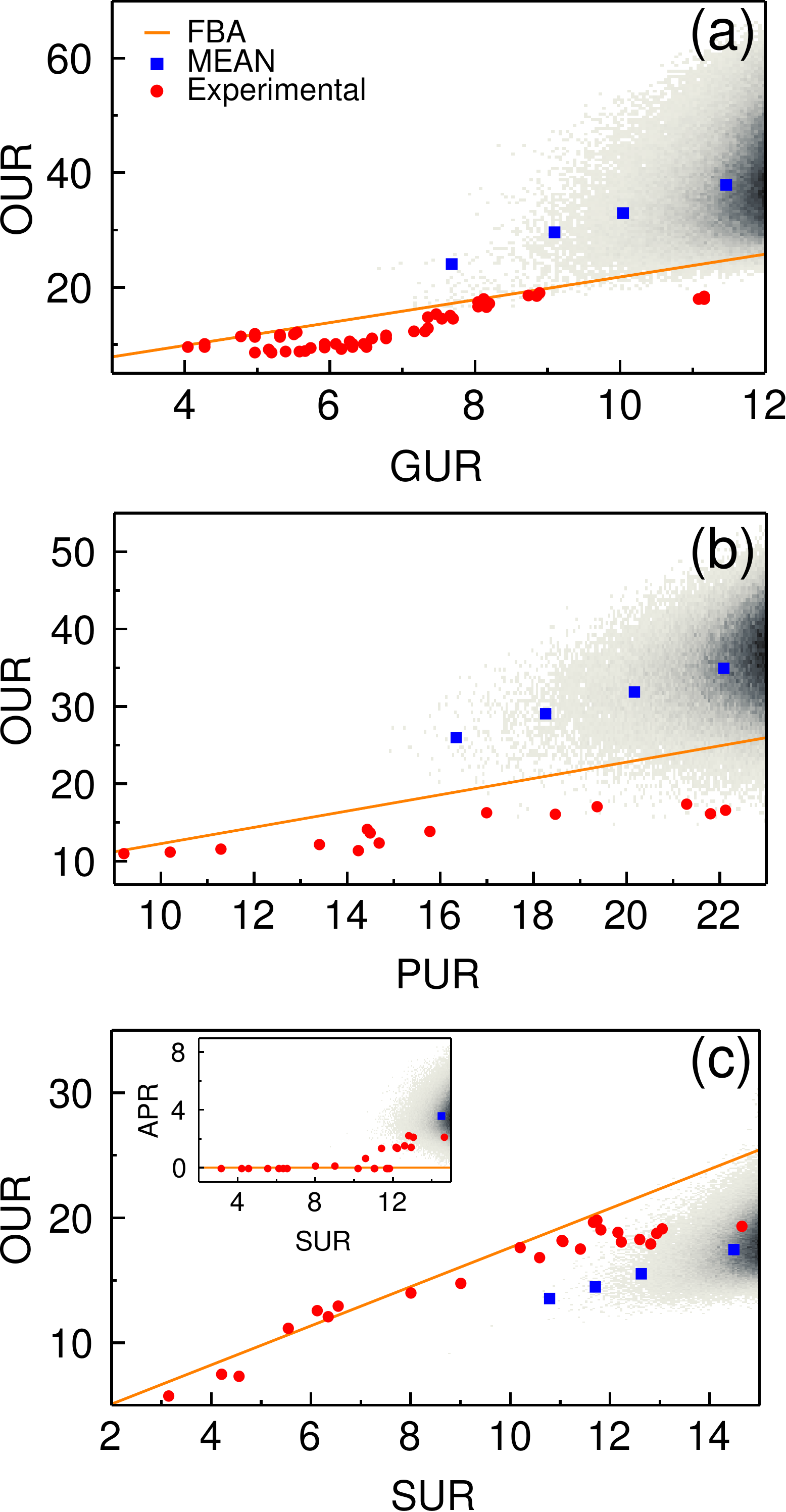}
\caption{ {\bf Comparison of predicted phenotypes and experimental data.} Sampled points in the FFP space with maximum carbon source upper bound are plotted in shaded grey, darkness is proportional to the number of points. Experimental data points are red circles. The {\it in silico}-defined line of optimality, representing FBA optimal growth solutions as a function of the upper bound uptake rate of the carbon source, is shown in orange. Blue squares correspond to FFP mean values for different carbon source upper bound uptake rates. {\bf (a)} Oxygen vs. glucose uptake rates, experimental data from~\cite{Ibarra:2002}. The FFP space is sampled with glucose bounded to $12$ mmol/(gDW$\cdot$h). {\bf (b)} Oxygen vs. pyruvate uptake rates, experimental data from~\cite{Fong:2003}. The FFP space is sampled with pyruvate bounded to $23$ mmol/(gDW$\cdot$h). {\bf (c)} Oxygen vs. succinate uptake rates, experimental data from~\cite{Edwards2001}. The FFP space is sampled with succinate bounded to $15$ mmol/(gDW$\cdot$h). {\em Inset} Acetate production rate vs. succinate uptake rate, experimental data from~\cite{Edwards2001}.}
\label{fig2}
\end{figure}

We focus on the relationship between primary carbon source uptakes and oxygen need to illustrate the potential of the FFP space as a benchmark to calibrate the deviation of {\it in silico} predicted optimal phenotypes from experimental observations. Sampled FFP states of {\it E. coli} core model, in particular FFP mean values as a function of the upper bound uptake rate of the carbon source, are compared with reported experimental data for oxygen uptakes in minimal medium with glucose, pyruvate, or succinate as a primary carbon source, Fig.~\ref{fig2}. We also included in the figures the line of optimality representing FBA optimal growth solutions. We used glucose experimental data points from~\cite{Ibarra:2002}, experimental results for pyruvate reported in~\cite{Fong:2003}, and experimental results in~\cite{Edwards2001} for the quantitative relationship between oxygen uptake rate and acetate production rate as a function of succinate uptake rate.  

In all cases, FBA-MBR reproduces well experimental data points in the low carbon source uptake region~\cite{Edwards2001}, where {\it E. coli} is indeed optimizing biomass yield. However, oxygen uptake rate saturates after some critical threshold of carbon source uptake rate, which depends on the carbon source, reaching a plateau which, among other possibilities, could be explained by the existence of a physiological enzymatic limit in oxygen uptake that lessens the capacity of the respiratory system~\cite{Varma:1993}. The plateau levels are $18.8 \pm 0.7$ mmol/(gDW$\cdot$h) for glucose~\cite{Edwards2001}, $16.8 \pm 0.4$ mmol/(gDW$\cdot$h) for pyruvate~\cite{Fong:2003}, and $19.49 \pm 0.78$ mmol/(gDW$\cdot$h) for succinate~\cite{Edwards2001}. In this region of high carbon source uptake, FBA-MBR predicts an oxygen uptake overestimated by around $25\%$ with respect to the values reported from experiments. While this amount is in principle large, the FFP space gives a standard that helps to calibrate it. 

We measured the eccentricity of experimental observations as their distance to the FFP mean. For glucose, this value is $19.4$, which makes the distance of $5.3$ between the FBA-MBR prediction and experimental data relatively low, Fig.~\ref{fig2}a. The distance of $8.2$ between the FBA-MBR prediction and experimental data is slightly worse for pyruvate, Fig.~\ref{fig2}b, in which case the eccentricity of experimental observations is of $18.4$. The disagreement between optimality predictions and experimental data is much more significative in the case of succinate, Fig.~\ref{fig2}c, for which the eccentricity of experimental observations is only of $4.3$, while the distance between the FBA-MBR prediction and experimental data is of $5.4$, meaning that the FFP mean is indeed more adjusted to observations. The case of acetate production for this carbon source is even more conspicuous, Fig.~\ref{fig2}c {\em Inset}. While FBA-MBR is still reproducing well the experimental results of no acetate production in the low succinate uptake region, it cannot predict production of acetate at any succinate uptake rate due to the fact that FBA-MBR in minimal medium with unlimited oxygen does not capture the enzymatic oxygen limitation. The FBA-MBR solution diverts resources to the production of ATP entirely through the oxidative phosphorylation pathway. Thus, it fails to reproduce experimental observations of acetate production in the region of high succinate uptake rates~\cite{Reiling:1985,El-Mansi:1989,Edwards2001,Wolfe:2005}. In contrast, most metabolic states in the FFP space are consistent with acetate production, so that in this case the FFP mean turns out as a good predictor of the experimentally observed metabolic behavior.

In summary, while FBA-MBR predictions seem accurate for low carbon source uptake rate states in minimal medium as seen previously~\cite{Edwards2001}, the experimental points diverge from the FBA-MBR prediction state when increased values of carbon source uptake rates are considered. Note that, in general, it is not straightforward to quantify the significance of the divergence. Here, we propose to use the FFP space as a reference standard. According to this calibration, we remarkably find that FBA optimal growth predictions of oxygen needs versus glucose, pyruvate, or succinate uptake are worse the more downstream the position of the carbon source into catalytic metabolism. Using the {\it E. coli} core metabolism, we have checked that the ratio of the maximum ATP production rate to the maximum oxygen uptake (both calculated by FBA optimization of ATP production rate) for the three carbon sources glucose, pyruvate, and succinate are respectively $2.9$, $2.6$, and $2.4$, so this ratio decreases as more downstream in the catalytic metabolism. These results are consistent with values reported in~\cite{Varma:1993}. FBA privileges energy production by diverting fluxes to oxidative phosphorylation providing maximum energy for growth, so that FBA should work worse the less effective the oxidation of the carbon source is for ATP synthesis. This can be explained in terms of departures of energy from substrate catabolism to functions other than growth, like basal maintenance, which become more relevant in relative terms as compared to the total energy production when the energy-to-redox ratios of the carbon substrate are lower \cite{Varman:2014}.

\subsection{High-biomass production FFP region displays aerobic fermentation in minimal medium with unlimited oxygen uptake}
We resampled the high-growth metabolic region of the {\it E. coli} core metabolism FFP space in glucose minimal medium with glucose upper bound of $10$ mmol/(gDW$\cdot$h), as in subsection 2a. We defined this region by setting a minimal threshold for the biomass production of $\ge 0.4$ mmol/(gDW$\cdot$h)~\cite{Vazquez:2008}, and produced a sampled with a final size of $10^5$ states. We note that phenotypes in this high-growth sample remain very close to the biomass yield threshold due to the exponential decrease of the number of feasible flux states with increased biomass production, as in the biomass flux profile in Fig.~\ref{fig1}a. 

In this region, we identified pathway utilization typical of proliferative microbial me\-ta\-bo\-lism, even when considering a minimal medium and unlimited oxygen uptake. This metabolic behavior is consistent with experimental data~\cite{Varma1994b,Edwards2001,Ibarra:2002} but it is unreachable by FBA models based on optimality principles (unless optimization is accompanied by auxiliary constraints not assumed in standard FBA implementations, like the solvent capacity constraint ~\cite{Vazquez:2008}, or by modelization beyond stoichiometric mass balance, introducing for instance thermodynamically feasible kinetics or enzyme synthesis~\cite{Molenaar:2009,Wortel:2014}). We checked that the by-products cannot be explained by FBA-MBR in minimal medium with unlimited oxygen supply since, in this optimization framework, metabolic fluxes are basically forced to ATP production through oxidative phosphorylation with excretion of CO$_2$ as waste. However, increasing the oxygen limitation in FBA-MBR results in secretion of formate, acetate, and ethanol --in that order--, with corresponding shifts in metabolic behavior~\cite{Varma:1993}. 

According to the FFP space of {\it E. coli} core metabolism, we observe that the high-biomass production FFP subsample is characterized by the secretion of small organic acid molecules, even when the supply of oxygen is unlimited. This fact points to the simultaneous utilization of glycolysis and oxidative phosphorylation to produce biomass and energy and so to suboptimal states. This observation is supported by results from $^{13}$C-metabolic flux analysis in {\it E. coli}~\cite{Chen:2011}, where repressed oxidative phosphorylation was proposed as responsible for the measured submaximal aerobic growth. Pathway utilization is illustrated in the schematic shown in Fig.~\ref{fig3}a. Quantitative relationships between the production of small organic acids molecules and glucose and oxygen uptake rates are shown in the remaining panels of Fig.~\ref{fig3}. Three-dimensional scatterplots for the production rates of formate, acetate, and ethanol are shown in Figs.~\ref{fig3}b,~\ref{fig3}d, and~\ref{fig3}f respectively, with projections into the three possible two-dimensional planes shown in Figs.~\ref{fig3}c,~\ref{fig3}e, and~\ref{fig3}g respectively. Figure S3 in SI gives results for lactate. As the levels of glucose and oxygen uptakes are raised, metabolic phenotypes can achieve an increased production of formate, acetate, and ethanol, even though the majority of feasible phenotypes remain at low organic acids production values. Due to the high-growth requirement, oxygen uptake is always high but its variability increases with glucose uptake increase around a value of approximately $41.2$ mmol/(gDW$\cdot$h), which clusters the majority of high-growth metabolic phenotypes. Interestingly, this oxygen uptake rate value marks a region in the FFP space with maximum potential production rates of formate, acetate, and ethanol. Above and below that value most states are concentrated in the range $[39,42]$ mmol/(gDW$\cdot$h).

\begin{figure}[htp]
\includegraphics[width=0.66\textwidth]{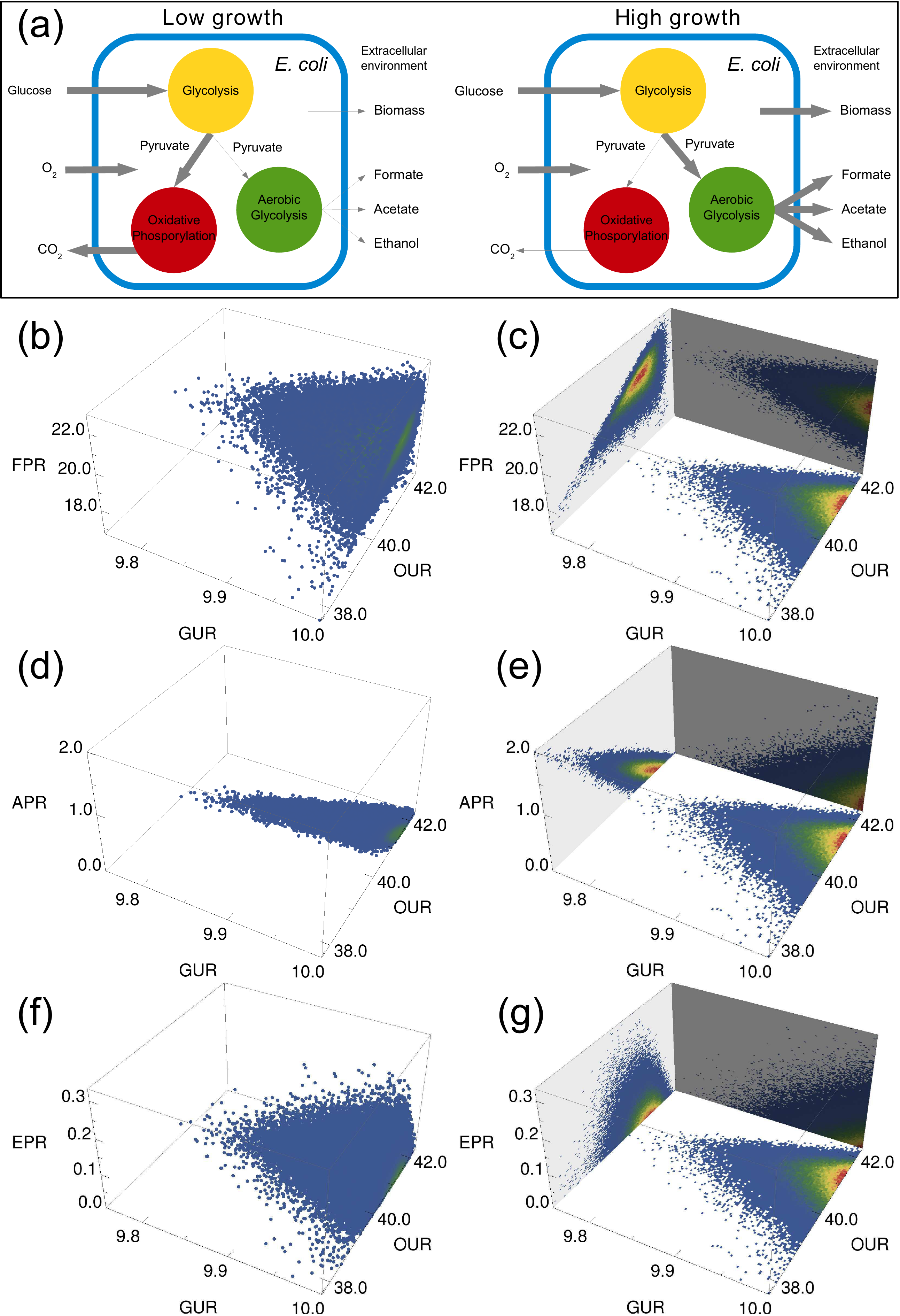}
\caption{{\bf High growth phenotypes of {\it E. coli} core metabolism on glucose minimal medium.} {\bf (a)} Schematic of pathway utilization in high-growth vs low-growth conditions. {\bf (b)} 3-dimensional scatterplot of formate production rate vs glucose and oxygen uptake rates. {\bf (c)} Density projections of (b) on each of the possible 2D planes, formate-glucose, formate-oxygen, and glucose-oxygen. {\bf (d)} 3-dimensional scatterplot of acetate production rate vs glucose and oxygen uptake rates. {\bf (e)} Density projections of (d) on each of the possible 2D planes, acetate-glucose, acetate-oxygen, and acetate-oxygen. {\bf (f)} 3-dimensional scatterplot of ethanol production rate vs glucose and oxygen uptake rates. {\bf (g)} Density projections of (f) on each of the possible 2D planes, ethanol-glucose, ethanol-oxygen, and glucose-oxygen.}
\label{fig3}
\end{figure} 

Taken together, these results indicate that, contrarily to FBA predictions, a high level of glucose uptake combined with unlimited oxygen can maintain the requirements of proliferative metabolism for biomass formation through aerobic fermentation even in minimal medium. At the same time, additional oxygen uptake diverts glucose back towards more efficient ATP production through oxidative phosphorylation. Hence, oxygen has the potential of regulating the glucose metabolic switch in which glucose uptake rates larger than a critical threshold around $5.0$ mmol/(gDW$\cdot$h)~\cite{Vazquez:2008} lead to a linearly increasing maximum organic acids by-products production by a gradual activation of aerobic fermentation and a slight decrease of oxidative phosphorylation. The reduction of glycolysis fluxes for oxygen sufficient conditions were reported previously in the context of $^{13}$C-metabolic flux analysis~\cite{Chen:2011}.

\section{Discussion}
The information content of the full FFP space of metabolic states in a certain environment provides us with an entire map to explore and evaluate metabolic behavior and capabilities. While optimality goals need to be tailored to conditions and produce singular optimal solutions that may not be consistent with experimental observations, we have nowadays sufficient computational and methodological capacity to produce and analyze full FFP maps. The latter offer a reference framework to put into perspective the likelihood of particular phenotypic states that, as shown, enables to uncover metabolic behaviors that are unreachable using models based on optimality principles. In fact, the location of metabolic flux distributions into precise optimal states has been challenged recently. Steady-state $^{13}$C-metabolic flux analysis applied to {\it E. coli}~\cite{Chen:2011} found submaximal growth in aerobic conditions. It has been proposed that metabolic flux evolves under the trade-off between two forces, optimality under one given condition and minimal adjustment between conditions~\cite{Schuetz:2012}. In this way, resilience to changing environments necessarily forces flux states to near-optimal but suboptimal regions of feasible flux states in order to maintain adaptability. 

In the FFP map of {\it E. coli} core metabolism in aerobic minimal medium, optimal growth states appear as eccentric and far from the bulk of more probable phenotypes represented by the FFP mean, which offers an ergodic perspective of the FFP space in which all states can be explored at random with equal probability. One of the uses of the method is precisely to evidence the effects of evolutionary pressure on organisms, which may actually result in eccentric flux states. On the other hand, the FFP space gives a standard to calibrate the deviation of optimal phenotypes from experimental observations. Oxygen consumption is a particularly interesting target for analysis since it has been identified as a trigger of metabolic shifts~\cite{Varma:1993,Losen:2004}. Interestingly, according to the FFP map as a reference standard, we found that in high-growth conditions FBA-MBR predictions of experimental observations for unlimited oxygen needs versus glucose, pyruvate, or succinate uptakes are worse the more downstream the uptake of the carbon source into the catalytic metabolic stream. This is consistent with the fact that the FBA-MBR solution diverts resources to the production of ATP entirely through the oxidative phosphorylation pathway, so that the more is the effective potential of the carbon source to recombine with oxygen to produce energy the more convergent will be the {\it in silico} prediction and the observed states. 

In order to correct FBA in high-growth conditions, some investigations restricted the solution space beyond mass balance and uptake bounds through additional thermodynamic, kinetic or physiological constraints, like the solvent capacity constraint quantifying the maximum amount of macromolecules that can occupy the intracellular space~\cite{Vazquez:2008}. Alternatively, the objective function implemented in FBA has been modified to nonlinear maximization of the ATP or biomass yield per flux unit~\cite{Schuetz:2007}. Other models consider constraints beyond stoichiometric mass balance, for instance thermodynamically feasible kinetics or enzyme synthesis~\cite{Molenaar:2009,Wortel:2014}. While these FBA modifications enhance some predictions, their effectiveness depends on the estimation of kinetic coefficients using empirical or experimental data. In contrast, the FFP map contains the set of all solutions determined solely on the basis of mass balance and upper and lower bounds for nutrients, and therefore it includes solutions compliant with physiological constraints or with the limitations imposed by complex metabolic regulation~\cite{Chubukov:2014}. In particular, the FFP space naturally displays all high-growth feasible states which show characteristic metabolic behaviors like aerobic fermentation with unlimited oxygen uptake even in minimal medium without the need to determine additional constants. This aerobic fermentation, apparently inefficient in terms of energy yield as compared to oxidative phosphorylation but demonstrated as a favorable catabolic state for all rapidly proliferating cells with high glucose uptake capacity~\cite{Vazquez:2008}, turns out as a probable metabolic phenotype even in minimal medium. Results in this direction have been reported using steady-state $^{13}$C-metabolic flux analysis, which has shown that {\it E. coli} grows suboptimally in glucose minimal media due to limited oxidative phosphorylation~\cite{Chen:2011}.

Beyond theoretical implications, FFP maps of microbial organisms can be of particular interest as tools for biotechnological applications, for instance in the engineering of {\it E. coli} fermentative metabolism as a fundamental cellular capacity for valuable industrial biocatalysis~\cite{Orencio-Trejo:2010}. In biomedicine, the investigation of FBA optimal phenotypes in the framework of the FFP map can help to contextualize disease phenotypes in comparison to normal states. For instance, FBA proved suitable for modelling complex diseases like cancer as it assumes that cancer cells maximize growth searching for metabolic flux distributions that produce essential biomass precursors at high rates~\cite{Price:2007,Folger:2011}. The analysis of the entire region of high-growth phenotypes will allow to reach and study a variety of suboptimal feasible flux states close to optimality but which cannot be reproduced by optimality principles, and so it opens new avenues for the understanding of general and fundamental mechanisms that characterize this disease across subtypes. 

\section*{Acknowledgments}
We thank Adam Palmer for pointing out to us Ref.~\cite{Schuetz:2007}

%\nolinenumbers

%\bibliography{ref}

\begin{thebibliography}{51}%
\makeatletter
\providecommand \@ifxundefined [1]{%
 \@ifx{#1\undefined}
}%
\providecommand \@ifnum [1]{%
 \ifnum #1\expandafter \@firstoftwo
 \else \expandafter \@secondoftwo
 \fi
}%
\providecommand \@ifx [1]{%
 \ifx #1\expandafter \@firstoftwo
 \else \expandafter \@secondoftwo
 \fi
}%
\providecommand \natexlab [1]{#1}%
\providecommand \enquote  [1]{``#1''}%
\providecommand \bibnamefont  [1]{#1}%
\providecommand \bibfnamefont [1]{#1}%
\providecommand \citenamefont [1]{#1}%
\providecommand \href@noop [0]{\@secondoftwo}%
\providecommand \href [0]{\begingroup \@sanitize@url \@href}%
\providecommand \@href[1]{\@@startlink{#1}\@@href}%
\providecommand \@@href[1]{\endgroup#1\@@endlink}%
\providecommand \@sanitize@url [0]{\catcode `\\12\catcode `\$12\catcode
  `\&12\catcode `\#12\catcode `\^12\catcode `\_12\catcode `\%12\relax}%
\providecommand \@@startlink[1]{}%
\providecommand \@@endlink[0]{}%
\providecommand \url  [0]{\begingroup\@sanitize@url \@url }%
\providecommand \@url [1]{\endgroup\@href {#1}{\urlprefix }}%
\providecommand \urlprefix  [0]{URL }%
\providecommand \Eprint [0]{\href }%
\providecommand \doibase [0]{http://dx.doi.org/}%
\providecommand \selectlanguage [0]{\@gobble}%
\providecommand \bibinfo  [0]{\@secondoftwo}%
\providecommand \bibfield  [0]{\@secondoftwo}%
\providecommand \translation [1]{[#1]}%
\providecommand \BibitemOpen [0]{}%
\providecommand \bibitemStop [0]{}%
\providecommand \bibitemNoStop [0]{.\EOS\space}%
\providecommand \EOS [0]{\spacefactor3000\relax}%
\providecommand \BibitemShut  [1]{\csname bibitem#1\endcsname}%
\let\auto@bib@innerbib\@empty
%</preamble>
\bibitem [{\citenamefont {Ibarra}\ \emph {et~al.}(2002)\citenamefont {Ibarra},
  \citenamefont {Edwards},\ and\ \citenamefont {Palsson}}]{Ibarra:2002}%
  \BibitemOpen
  \bibfield  {author} {\bibinfo {author} {\bibfnamefont {R.~U.}\ \bibnamefont
  {Ibarra}}, \bibinfo {author} {\bibfnamefont {J.~S.}\ \bibnamefont {Edwards}},
  \ and\ \bibinfo {author} {\bibfnamefont {B.~\O.}\ \bibnamefont {Palsson}},\
  }\bibfield  {title} {\enquote {\bibinfo {title} {{{\it Escherichia coli} K-12
  undergoes adaptive evolution to achieve {\it in silico} predicted optimal
  growth}},}\ }\href@noop {} {\bibfield  {journal} {\bibinfo  {journal}
  {Nature}\ }\textbf {\bibinfo {volume} {420}},\ \bibinfo {pages} {186--189}
  (\bibinfo {year} {2002})}\BibitemShut {NoStop}%
\bibitem [{\citenamefont {Blank}\ \emph {et~al.}(2005)\citenamefont {Blank},
  \citenamefont {Kuepfer},\ and\ \citenamefont {Sauer}}]{blank2005}%
  \BibitemOpen
  \bibfield  {author} {\bibinfo {author} {\bibfnamefont {L.~M.}\ \bibnamefont
  {Blank}}, \bibinfo {author} {\bibfnamefont {L.}~\bibnamefont {Kuepfer}}, \
  and\ \bibinfo {author} {\bibfnamefont {U.}~\bibnamefont {Sauer}},\ }\bibfield
   {title} {\enquote {\bibinfo {title} {{Large-scale $^{13}$C-flux analysis
  reveals mechanistic principles of metabolic network robustness to null
  mutations in yeast}},}\ }\href@noop {} {\bibfield  {journal} {\bibinfo
  {journal} {Genome Biology}\ }\textbf {\bibinfo {volume} {6}},\ \bibinfo
  {pages} {R49} (\bibinfo {year} {2005})}\BibitemShut {NoStop}%
\bibitem [{\citenamefont {Frick}\ and\ \citenamefont
  {Whittmann}(2005)}]{Frick:2005}%
  \BibitemOpen
  \bibfield  {author} {\bibinfo {author} {\bibfnamefont {O.}~\bibnamefont
  {Frick}}\ and\ \bibinfo {author} {\bibfnamefont {C.}~\bibnamefont
  {Whittmann}},\ }\bibfield  {title} {\enquote {\bibinfo {title}
  {{Characterization of the metabolic shift between oxidative and fermentative
  growth in {\it Saccharomyces cerevisiae} by comparative $^{13}$C flux
  analysis}},}\ }\href@noop {} {\bibfield  {journal} {\bibinfo  {journal}
  {Microbial Cell Factories}\ }\textbf {\bibinfo {volume} {4}},\ \bibinfo
  {pages} {30} (\bibinfo {year} {2005})}\BibitemShut {NoStop}%
\bibitem [{\citenamefont {Deken}(1996)}]{Deken:1996}%
  \BibitemOpen
  \bibfield  {author} {\bibinfo {author} {\bibfnamefont {R.~H.~De}\
  \bibnamefont {Deken}},\ }\bibfield  {title} {\enquote {\bibinfo {title} {The
  crabtree effect: a regulatory system in yeast},}\ }\href@noop {} {\bibfield
  {journal} {\bibinfo  {journal} {Journal of General Microbiology}\ }\textbf
  {\bibinfo {volume} {44}},\ \bibinfo {pages} {149--156} (\bibinfo {year}
  {1996})}\BibitemShut {NoStop}%
\bibitem [{\citenamefont {Paczia}\ \emph {et~al.}(2012)\citenamefont {Paczia},
  \citenamefont {Nilgen}, \citenamefont {Lehmann}, \citenamefont {G{\"a}tgens},
  \citenamefont {Wiechert},\ and\ \citenamefont {Noack}}]{Paczia:2012}%
  \BibitemOpen
  \bibfield  {author} {\bibinfo {author} {\bibfnamefont {Nicole}\ \bibnamefont
  {Paczia}}, \bibinfo {author} {\bibfnamefont {Anke}\ \bibnamefont {Nilgen}},
  \bibinfo {author} {\bibfnamefont {Tobias}\ \bibnamefont {Lehmann}}, \bibinfo
  {author} {\bibfnamefont {Jochem}\ \bibnamefont {G{\"a}tgens}}, \bibinfo
  {author} {\bibfnamefont {Wolfgang}\ \bibnamefont {Wiechert}}, \ and\ \bibinfo
  {author} {\bibfnamefont {Stephan}\ \bibnamefont {Noack}},\ }\bibfield
  {title} {\enquote {\bibinfo {title} {Extensive exometabolome analysis reveals
  extended overflow metabolism in various microorganisms},}\ }\href@noop {}
  {\bibfield  {journal} {\bibinfo  {journal} {Microbial Cell Factories}\
  }\textbf {\bibinfo {volume} {11}},\ \bibinfo {pages} {122} (\bibinfo {year}
  {2012})}\BibitemShut {NoStop}%
\bibitem [{\citenamefont {Heiden}\ \emph {et~al.}(2009)\citenamefont {Heiden},
  \citenamefont {Cantley},\ and\ \citenamefont {Thompson}}]{VanderHeiden:2009}%
  \BibitemOpen
  \bibfield  {author} {\bibinfo {author} {\bibfnamefont {M.~G.~Vander}\
  \bibnamefont {Heiden}}, \bibinfo {author} {\bibfnamefont {L.~C.}\
  \bibnamefont {Cantley}}, \ and\ \bibinfo {author} {\bibfnamefont {C.~B.}\
  \bibnamefont {Thompson}},\ }\bibfield  {title} {\enquote {\bibinfo {title}
  {{Understanding the Warburg Effect: The Metabolic Requirements of Cell
  Proliferation}},}\ }\href@noop {} {\bibfield  {journal} {\bibinfo  {journal}
  {Science}\ }\textbf {\bibinfo {volume} {324}},\ \bibinfo {pages} {1029--1033}
  (\bibinfo {year} {2009})}\BibitemShut {NoStop}%
\bibitem [{\citenamefont {Orth}\ \emph
  {et~al.}(2010{\natexlab{a}})\citenamefont {Orth}, \citenamefont {Thiele},\
  and\ \citenamefont {Palsson}}]{Orth:2010}%
  \BibitemOpen
  \bibfield  {author} {\bibinfo {author} {\bibfnamefont {J.~D.}\ \bibnamefont
  {Orth}}, \bibinfo {author} {\bibfnamefont {I.}~\bibnamefont {Thiele}}, \ and\
  \bibinfo {author} {\bibfnamefont {B.~{\O.}}\ \bibnamefont {Palsson}},\
  }\bibfield  {title} {\enquote {\bibinfo {title} {What is flux balance
  analysis?}}\ }\href@noop {} {\bibfield  {journal} {\bibinfo  {journal}
  {Nature Biotechnology}\ }\textbf {\bibinfo {volume} {28}},\ \bibinfo {pages}
  {245--248} (\bibinfo {year} {2010}{\natexlab{a}})}\BibitemShut {NoStop}%
\bibitem [{\citenamefont {Overbeek}\ \emph {et~al.}(2005)\citenamefont
  {Overbeek} \emph {et~al.}}]{overbeek05}%
  \BibitemOpen
  \bibfield  {author} {\bibinfo {author} {\bibfnamefont {R.}~\bibnamefont
  {Overbeek}} \emph {et~al.},\ }\bibfield  {title} {\enquote {\bibinfo {title}
  {The subsystems approach to genome annotation and its use in the project to
  annotate 1000 genomes},}\ }\href@noop {} {\bibfield  {journal} {\bibinfo
  {journal} {Nucleic Acids Research}\ }\textbf {\bibinfo {volume} {33}},\
  \bibinfo {pages} {5691--5702} (\bibinfo {year} {2005})}\BibitemShut {NoStop}%
\bibitem [{\citenamefont {Blattner}\ \emph {et~al.}(1997)\citenamefont
  {Blattner} \emph {et~al.}}]{Blattner1997}%
  \BibitemOpen
  \bibfield  {author} {\bibinfo {author} {\bibfnamefont {F.~R.}\ \bibnamefont
  {Blattner}} \emph {et~al.},\ }\bibfield  {title} {\enquote {\bibinfo {title}
  {{The complete genome sequence of {\it Escherichia coli} K-12}},}\
  }\href@noop {} {\bibfield  {journal} {\bibinfo  {journal} {Science}\ }\textbf
  {\bibinfo {volume} {277}},\ \bibinfo {pages} {1453--1474} (\bibinfo {year}
  {1997})}\BibitemShut {NoStop}%
\bibitem [{\citenamefont {Feist}\ \emph {et~al.}(2007)\citenamefont {Feist}
  \emph {et~al.}}]{Feist:2007}%
  \BibitemOpen
  \bibfield  {author} {\bibinfo {author} {\bibfnamefont {A.}~\bibnamefont
  {Feist}} \emph {et~al.},\ }\bibfield  {title} {\enquote {\bibinfo {title} {{A
  genome-scale metabolic reconstruction for {\it Escherichia coli} K-12 MG1655
  that accounts for 1260 ORFs and thermodynamic information}},}\ }\href@noop {}
  {\bibfield  {journal} {\bibinfo  {journal} {Molecular Systems Biology}\
  }\textbf {\bibinfo {volume} {3}},\ \bibinfo {pages} {121} (\bibinfo {year}
  {2007})}\BibitemShut {NoStop}%
\bibitem [{\citenamefont {Oh}\ \emph {et~al.}(2007)\citenamefont {Oh} \emph
  {et~al.}}]{oh07}%
  \BibitemOpen
  \bibfield  {author} {\bibinfo {author} {\bibfnamefont {Y.~K.}\ \bibnamefont
  {Oh}} \emph {et~al.},\ }\bibfield  {title} {\enquote {\bibinfo {title}
  {Genome-scale reconstruction of metabolic network in {\it bacillus subtilis}
  based on high-throughput phenotyping and gene essentiality data.}}\
  }\href@noop {} {\bibfield  {journal} {\bibinfo  {journal} {Journal of
  Biological Chemistry}\ }\textbf {\bibinfo {volume} {282}},\ \bibinfo {pages}
  {28791--28799} (\bibinfo {year} {2007})}\BibitemShut {NoStop}%
\bibitem [{\citenamefont {Mo}\ \emph {et~al.}(2009)\citenamefont {Mo},
  \citenamefont {Palsson},\ and\ \citenamefont {Herrg{\aa}rd}}]{mo09}%
  \BibitemOpen
  \bibfield  {author} {\bibinfo {author} {\bibfnamefont {M.~L.}\ \bibnamefont
  {Mo}}, \bibinfo {author} {\bibfnamefont {B.~{\O}.}\ \bibnamefont {Palsson}},
  \ and\ \bibinfo {author} {\bibfnamefont {M.~J.}\ \bibnamefont
  {Herrg{\aa}rd}},\ }\bibfield  {title} {\enquote {\bibinfo {title} {Connecting
  extracellular metabolomic measurements to intracellular flux states in
  yeast.}}\ }\href@noop {} {\bibfield  {journal} {\bibinfo  {journal} {BMC
  Systems Biology}\ }\textbf {\bibinfo {volume} {3}},\ \bibinfo {pages} {37}
  (\bibinfo {year} {2009})}\BibitemShut {NoStop}%
\bibitem [{\citenamefont {Almaas}\ \emph {et~al.}(2005)\citenamefont {Almaas},
  \citenamefont {Oltvai},\ and\ \citenamefont {Barab\'asi}}]{Almaas:2005}%
  \BibitemOpen
  \bibfield  {author} {\bibinfo {author} {\bibfnamefont {E.}~\bibnamefont
  {Almaas}}, \bibinfo {author} {\bibfnamefont {Z.~N.}\ \bibnamefont {Oltvai}},
  \ and\ \bibinfo {author} {\bibfnamefont {A.~L.}\ \bibnamefont {Barab\'asi}},\
  }\bibfield  {title} {\enquote {\bibinfo {title} {The activity reaction core
  and plasticity of metabolic networks},}\ }\href@noop {} {\bibfield  {journal}
  {\bibinfo  {journal} {PLoS Computational Biology}\ }\textbf {\bibinfo
  {volume} {1}},\ \bibinfo {pages} {0557--0563} (\bibinfo {year}
  {2005})}\BibitemShut {NoStop}%
\bibitem [{\citenamefont {Suthers}\ \emph {et~al.}(2009)\citenamefont
  {Suthers}, \citenamefont {Zomorrodi},\ and\ \citenamefont
  {Maranas}}]{Suthers:2009}%
  \BibitemOpen
  \bibfield  {author} {\bibinfo {author} {\bibfnamefont {P.~F.}\ \bibnamefont
  {Suthers}}, \bibinfo {author} {\bibfnamefont {A.}~\bibnamefont {Zomorrodi}},
  \ and\ \bibinfo {author} {\bibfnamefont {C.~D.}\ \bibnamefont {Maranas}},\
  }\bibfield  {title} {\enquote {\bibinfo {title} {Genome-scale gene/reaction
  essentiality and synthetic lethality analysis},}\ }\href@noop {} {\bibfield
  {journal} {\bibinfo  {journal} {Molecular Systems Biology}\ }\textbf
  {\bibinfo {volume} {5}},\ \bibinfo {pages} {301} (\bibinfo {year}
  {2009})}\BibitemShut {NoStop}%
\bibitem [{\citenamefont {Segr\`{e}}\ \emph {et~al.}(2002)\citenamefont
  {Segr\`{e}}, \citenamefont {Vitkup},\ and\ \citenamefont
  {Church}}]{Segre:2002}%
  \BibitemOpen
  \bibfield  {author} {\bibinfo {author} {\bibfnamefont {D.}~\bibnamefont
  {Segr\`{e}}}, \bibinfo {author} {\bibfnamefont {D.}~\bibnamefont {Vitkup}}, \
  and\ \bibinfo {author} {\bibfnamefont {G.~M.}\ \bibnamefont {Church}},\
  }\bibfield  {title} {\enquote {\bibinfo {title} {Analysis of optimality in
  natural and perturbed metabolic networks},}\ }\href@noop {} {\bibfield
  {journal} {\bibinfo  {journal} {Proc. Natl. Acad. Sci. USA}\ }\textbf
  {\bibinfo {volume} {99}},\ \bibinfo {pages} {15112--15117} (\bibinfo {year}
  {2002})}\BibitemShut {NoStop}%
\bibitem [{\citenamefont {Schuetz}\ \emph {et~al.}(2007)\citenamefont
  {Schuetz}, \citenamefont {Kuepfer},\ and\ \citenamefont
  {Sauer}}]{Schuetz:2007}%
  \BibitemOpen
  \bibfield  {author} {\bibinfo {author} {\bibfnamefont {R.}~\bibnamefont
  {Schuetz}}, \bibinfo {author} {\bibfnamefont {L.}~\bibnamefont {Kuepfer}}, \
  and\ \bibinfo {author} {\bibfnamefont {U.}~\bibnamefont {Sauer}},\ }\bibfield
   {title} {\enquote {\bibinfo {title} {{Systematic evaluation of objective
  functions for predicting intracellular fluxes in {\it Escherichia coli}}},}\
  }\href@noop {} {\bibfield  {journal} {\bibinfo  {journal} {Molecular Systems
  Biology}\ }\textbf {\bibinfo {volume} {3}},\ \bibinfo {pages} {119} (\bibinfo
  {year} {2007})}\BibitemShut {NoStop}%
\bibitem [{\citenamefont {Schuetz}\ \emph {et~al.}(2012)\citenamefont
  {Schuetz}, \citenamefont {Zamboni}, \citenamefont {Zampieri}, \citenamefont
  {Heinemann},\ and\ \citenamefont {Sauer}}]{Schuetz:2012}%
  \BibitemOpen
  \bibfield  {author} {\bibinfo {author} {\bibfnamefont {R.}~\bibnamefont
  {Schuetz}}, \bibinfo {author} {\bibfnamefont {N.}~\bibnamefont {Zamboni}},
  \bibinfo {author} {\bibfnamefont {M.}~\bibnamefont {Zampieri}}, \bibinfo
  {author} {\bibfnamefont {M.}~\bibnamefont {Heinemann}}, \ and\ \bibinfo
  {author} {\bibfnamefont {U.}~\bibnamefont {Sauer}},\ }\bibfield  {title}
  {\enquote {\bibinfo {title} {Multidimensional optimality of microbial
  metabolism},}\ }\href@noop {} {\bibfield  {journal} {\bibinfo  {journal}
  {Science}\ }\textbf {\bibinfo {volume} {336}},\ \bibinfo {pages} {601--604}
  (\bibinfo {year} {2012})}\BibitemShut {NoStop}%
\bibitem [{\citenamefont {Schuster}\ and\ \citenamefont
  {Hilgetag}(1994)}]{Schuster:1994}%
  \BibitemOpen
  \bibfield  {author} {\bibinfo {author} {\bibfnamefont {S.}~\bibnamefont
  {Schuster}}\ and\ \bibinfo {author} {\bibfnamefont {C.}~\bibnamefont
  {Hilgetag}},\ }\bibfield  {title} {\enquote {\bibinfo {title} {On elementary
  flux modes in biochemical reaction systems at steady state},}\ }\href@noop {}
  {\bibfield  {journal} {\bibinfo  {journal} {J. Biol. Syst.}\ }\textbf
  {\bibinfo {volume} {2}},\ \bibinfo {pages} {165--182} (\bibinfo {year}
  {1994})}\BibitemShut {NoStop}%
\bibitem [{\citenamefont {Schuster}\ \emph {et~al.}(2000)\citenamefont
  {Schuster}, \citenamefont {Fell},\ and\ \citenamefont
  {Dandekar}}]{Schuster:2000}%
  \BibitemOpen
  \bibfield  {author} {\bibinfo {author} {\bibfnamefont {S.}~\bibnamefont
  {Schuster}}, \bibinfo {author} {\bibfnamefont {D.~A.}\ \bibnamefont {Fell}},
  \ and\ \bibinfo {author} {\bibfnamefont {T.}~\bibnamefont {Dandekar}},\
  }\bibfield  {title} {\enquote {\bibinfo {title} {A general definition of
  metabolic pathways useful for systematic organization and analysis of complex
  metabolic networks},}\ }\href@noop {} {\bibfield  {journal} {\bibinfo
  {journal} {Nat. Biotechnol.}\ }\textbf {\bibinfo {volume} {18}},\ \bibinfo
  {pages} {326--332} (\bibinfo {year} {2000})}\BibitemShut {NoStop}%
\bibitem [{\citenamefont {Acu{\~n}a}\ \emph {et~al.}(2010)\citenamefont
  {Acu{\~n}a}, \citenamefont {Marchetti-Spaccamela}, \citenamefont {Sagot},\
  and\ \citenamefont {Stougie}}]{Acuna:2010}%
  \BibitemOpen
  \bibfield  {author} {\bibinfo {author} {\bibfnamefont {V.}~\bibnamefont
  {Acu{\~n}a}}, \bibinfo {author} {\bibfnamefont {A.}~\bibnamefont
  {Marchetti-Spaccamela}}, \bibinfo {author} {\bibfnamefont {M.~F.}\
  \bibnamefont {Sagot}}, \ and\ \bibinfo {author} {\bibfnamefont
  {L.}~\bibnamefont {Stougie}},\ }\bibfield  {title} {\enquote {\bibinfo
  {title} {A note on the complexity of finding and enumerating elementary
  modes},}\ }\href@noop {} {\bibfield  {journal} {\bibinfo  {journal}
  {Biosystems}\ ,\ \bibinfo {pages} {210--214}} (\bibinfo {year}
  {2010})}\BibitemShut {NoStop}%
\bibitem [{\citenamefont {Orth}\ \emph
  {et~al.}(2010{\natexlab{b}})\citenamefont {Orth}, \citenamefont {Fleming},\
  and\ \citenamefont {Palsson}}]{orth2009}%
  \BibitemOpen
  \bibfield  {author} {\bibinfo {author} {\bibfnamefont {J.~D.}\ \bibnamefont
  {Orth}}, \bibinfo {author} {\bibfnamefont {R.~M.}\ \bibnamefont {Fleming}}, \
  and\ \bibinfo {author} {\bibfnamefont {B.~{\O}.}\ \bibnamefont {Palsson}},\
  }\bibfield  {title} {\enquote {\bibinfo {title} {{Reconstruction and use of
  microbial metabolic networks: the core Escherichia coli metabolic model as an
  educational guide}},}\ }\href@noop {} {\bibfield  {journal} {\bibinfo
  {journal} {{\it Escherichia coli} and {\it Salmonella}: Cellular and
  Molecular Biology}\ }\textbf {\bibinfo {volume} {10.2.1}} (\bibinfo {year}
  {2010}{\natexlab{b}})}\BibitemShut {NoStop}%
\bibitem [{\citenamefont {Jungreuthmayer}\ and\ \citenamefont
  {Zanghellini}(2012)}]{Jungreuthmayer:2012}%
  \BibitemOpen
  \bibfield  {author} {\bibinfo {author} {\bibfnamefont {C.}~\bibnamefont
  {Jungreuthmayer}}\ and\ \bibinfo {author} {\bibfnamefont {J.}~\bibnamefont
  {Zanghellini}},\ }\bibfield  {title} {\enquote {\bibinfo {title} {Designing
  optimal cell factories: Integer programing couples elementary mode analysis
  with regulation},}\ }\href@noop {} {\bibfield  {journal} {\bibinfo  {journal}
  {BMC Syst. Biol.}\ }\textbf {\bibinfo {volume} {6}},\ \bibinfo {pages} {103}
  (\bibinfo {year} {2012})}\BibitemShut {NoStop}%
\bibitem [{\citenamefont {Gebauer}\ \emph {et~al.}(2012)\citenamefont
  {Gebauer}, \citenamefont {Schuster}, \citenamefont {de~Figueiredo},\ and\
  \citenamefont {Kaleta}}]{Gebauer:2012}%
  \BibitemOpen
  \bibfield  {author} {\bibinfo {author} {\bibfnamefont {J.}~\bibnamefont
  {Gebauer}}, \bibinfo {author} {\bibfnamefont {S.}~\bibnamefont {Schuster}},
  \bibinfo {author} {\bibfnamefont {L.~F.}\ \bibnamefont {de~Figueiredo}}, \
  and\ \bibinfo {author} {\bibfnamefont {C.}~\bibnamefont {Kaleta}},\
  }\bibfield  {title} {\enquote {\bibinfo {title} {Detecting and investigating
  substrate cycles in a genome-scale human metabolic network},}\ }\href@noop {}
  {\bibfield  {journal} {\bibinfo  {journal} {FEBS J.}\ }\textbf {\bibinfo
  {volume} {279}},\ \bibinfo {pages} {3192--3202} (\bibinfo {year}
  {2012})}\BibitemShut {NoStop}%
\bibitem [{\citenamefont {Schilling}\ \emph {et~al.}(2000)\citenamefont
  {Schilling}, \citenamefont {Edwards}, \citenamefont {Letscher},\ and\
  \citenamefont {Palsson}}]{Schilling:2000}%
  \BibitemOpen
  \bibfield  {author} {\bibinfo {author} {\bibfnamefont {C.~H.}\ \bibnamefont
  {Schilling}}, \bibinfo {author} {\bibfnamefont {J.~S.}\ \bibnamefont
  {Edwards}}, \bibinfo {author} {\bibfnamefont {D.}~\bibnamefont {Letscher}}, \
  and\ \bibinfo {author} {\bibfnamefont {B.~\O.}\ \bibnamefont {Palsson}},\
  }\bibfield  {title} {\enquote {\bibinfo {title} {Combining pathway analysis
  with flux balance analysis for the comprehensive study of metabolic
  systems},}\ }\href@noop {} {\bibfield  {journal} {\bibinfo  {journal}
  {Biotechnology and Bioengineering}\ }\textbf {\bibinfo {volume} {71}},\
  \bibinfo {pages} {286--306} (\bibinfo {year} {2000})}\BibitemShut {NoStop}%
\bibitem [{\citenamefont {Price}\ \emph {et~al.}(2004)\citenamefont {Price},
  \citenamefont {Schellenberger},\ and\ \citenamefont {Palsson}}]{Price:2004}%
  \BibitemOpen
  \bibfield  {author} {\bibinfo {author} {\bibfnamefont {N.~D.}\ \bibnamefont
  {Price}}, \bibinfo {author} {\bibfnamefont {J.}~\bibnamefont
  {Schellenberger}}, \ and\ \bibinfo {author} {\bibfnamefont {B.~\O.}\
  \bibnamefont {Palsson}},\ }\bibfield  {title} {\enquote {\bibinfo {title}
  {Uniform sampling of steady-state flux spaces: Means to design experiments
  and to interpret enzymopathies},}\ }\href@noop {} {\bibfield  {journal}
  {\bibinfo  {journal} {Biophysical Journal}\ }\textbf {\bibinfo {volume}
  {87}},\ \bibinfo {pages} {2172--2186} (\bibinfo {year} {2004})}\BibitemShut
  {NoStop}%
\bibitem [{\citenamefont {Wiback}\ \emph {et~al.}(2004)\citenamefont {Wiback},
  \citenamefont {Famili}, \citenamefont {Greenberg},\ and\ \citenamefont
  {Palsson}}]{Wiback:2004}%
  \BibitemOpen
  \bibfield  {author} {\bibinfo {author} {\bibfnamefont {S.~J.}\ \bibnamefont
  {Wiback}}, \bibinfo {author} {\bibfnamefont {I.}~\bibnamefont {Famili}},
  \bibinfo {author} {\bibfnamefont {H.~J.}\ \bibnamefont {Greenberg}}, \ and\
  \bibinfo {author} {\bibfnamefont {B.~\O.}\ \bibnamefont {Palsson}},\
  }\bibfield  {title} {\enquote {\bibinfo {title} {Monte carlo sampling can be
  used to determine the size and shape of the steady-state flux space},}\
  }\href@noop {} {\bibfield  {journal} {\bibinfo  {journal} {Journal of
  Theoretical Biology}\ }\textbf {\bibinfo {volume} {228}},\ \bibinfo {pages}
  {437--447} (\bibinfo {year} {2004})}\BibitemShut {NoStop}%
\bibitem [{\citenamefont {Almaas}\ \emph {et~al.}(2004)\citenamefont {Almaas},
  \citenamefont {Kov\'{a}cs}, \citenamefont {Vicsek}, \citenamefont {Oltvai},\
  and\ \citenamefont {Barab{\'a}si}}]{Almaas:2004}%
  \BibitemOpen
  \bibfield  {author} {\bibinfo {author} {\bibfnamefont {E.}~\bibnamefont
  {Almaas}}, \bibinfo {author} {\bibfnamefont {B.}~\bibnamefont {Kov\'{a}cs}},
  \bibinfo {author} {\bibfnamefont {T.}~\bibnamefont {Vicsek}}, \bibinfo
  {author} {\bibfnamefont {Z.~N.}\ \bibnamefont {Oltvai}}, \ and\ \bibinfo
  {author} {\bibfnamefont {Albert-L{\'a}szl{\'o}}\ \bibnamefont
  {Barab{\'a}si}},\ }\bibfield  {title} {\enquote {\bibinfo {title} {{Global
  Organization of metabolic fluxes in the bacterium {\it Escherichia coli}}},}\
  }\href@noop {} {\bibfield  {journal} {\bibinfo  {journal} {Nature}\ }\textbf
  {\bibinfo {volume} {427}},\ \bibinfo {pages} {839--843} (\bibinfo {year}
  {2004})}\BibitemShut {NoStop}%
\bibitem [{\citenamefont {Massucci}\ \emph {et~al.}(2013)\citenamefont
  {Massucci}, \citenamefont {Font-Clos}, \citenamefont {Martino},\ and\
  \citenamefont {Castillo}}]{Massucci:2013}%
  \BibitemOpen
  \bibfield  {author} {\bibinfo {author} {\bibfnamefont {F.~A.}\ \bibnamefont
  {Massucci}}, \bibinfo {author} {\bibfnamefont {F.}~\bibnamefont {Font-Clos}},
  \bibinfo {author} {\bibfnamefont {A.~De}\ \bibnamefont {Martino}}, \ and\
  \bibinfo {author} {\bibfnamefont {I.~P\'{e}rez}\ \bibnamefont {Castillo}},\
  }\bibfield  {title} {\enquote {\bibinfo {title} {A novel methodology to
  estimate metabolic flux distributions in constraint-based models},}\
  }\href@noop {} {\bibfield  {journal} {\bibinfo  {journal} {Metabolites}\
  }\textbf {\bibinfo {volume} {3}},\ \bibinfo {pages} {838--852} (\bibinfo
  {year} {2013})}\BibitemShut {NoStop}%
\bibitem [{\citenamefont {Krauth}\ and\ \citenamefont
  {Mezard}(1987)}]{Krauth:1987}%
  \BibitemOpen
  \bibfield  {author} {\bibinfo {author} {\bibfnamefont {W.}~\bibnamefont
  {Krauth}}\ and\ \bibinfo {author} {\bibfnamefont {M.}~\bibnamefont
  {Mezard}},\ }\bibfield  {title} {\enquote {\bibinfo {title} {Learning
  algorithms with optimal stability in neural networks},}\ }\href@noop {}
  {\bibfield  {journal} {\bibinfo  {journal} {Journal of Physics A}\ }\textbf
  {\bibinfo {volume} {20}},\ \bibinfo {pages} {L745--L752} (\bibinfo {year}
  {1987})}\BibitemShut {NoStop}%
\bibitem [{\citenamefont {Lov{\'a}sz}(1999)}]{Lovasz1999}%
  \BibitemOpen
  \bibfield  {author} {\bibinfo {author} {\bibfnamefont {L.}~\bibnamefont
  {Lov{\'a}sz}},\ }\bibfield  {title} {\enquote {\bibinfo {title} {{Hit-and-Run
  mixes fast}},}\ }\href@noop {} {\bibfield  {journal} {\bibinfo  {journal}
  {Mathematical Programming}\ }\textbf {\bibinfo {volume} {86}},\ \bibinfo
  {pages} {443--461} (\bibinfo {year} {1999})}\BibitemShut {NoStop}%
\bibitem [{\citenamefont {Pearson}(1901)}]{Pearson1901}%
  \BibitemOpen
  \bibfield  {author} {\bibinfo {author} {\bibfnamefont {K.}~\bibnamefont
  {Pearson}},\ }\bibfield  {title} {\enquote {\bibinfo {title} {{LIII. On lines
  and planes of closest fit to systems of points in space}},}\ }\href@noop {}
  {\bibfield  {journal} {\bibinfo  {journal} {Philosophical Magazine Series 6}\
  }\textbf {\bibinfo {volume} {2}},\ \bibinfo {pages} {559--572} (\bibinfo
  {year} {1901})}\BibitemShut {NoStop}%
\bibitem [{\citenamefont {Jolliffe}(2002)}]{Jolliffe2002}%
  \BibitemOpen
  \bibfield  {author} {\bibinfo {author} {\bibfnamefont {I.}~\bibnamefont
  {Jolliffe}},\ }\href@noop {} {\emph {\bibinfo {title} {Principal Component
  Analysis}}}\ (\bibinfo  {publisher} {Springer},\ \bibinfo {address} {New
  York, USA},\ \bibinfo {year} {2002})\BibitemShut {NoStop}%
\bibitem [{\citenamefont {Barrett}\ \emph {et~al.}(2009)\citenamefont
  {Barrett}, \citenamefont {Herrgard},\ and\ \citenamefont
  {Palsson}}]{Barrett:2009}%
  \BibitemOpen
  \bibfield  {author} {\bibinfo {author} {\bibfnamefont {C.~L.}\ \bibnamefont
  {Barrett}}, \bibinfo {author} {\bibfnamefont {M.~J.}\ \bibnamefont
  {Herrgard}}, \ and\ \bibinfo {author} {\bibfnamefont {B.~{\O.}}\ \bibnamefont
  {Palsson}},\ }\bibfield  {title} {\enquote {\bibinfo {title} {Decomposing
  complex reaction networks using random sampling, principal component analysis
  and basis rotation},}\ }\href@noop {} {\bibfield  {journal} {\bibinfo
  {journal} {BMC Systems Biology}\ }\textbf {\bibinfo {volume} {3}},\ \bibinfo
  {pages} {30} (\bibinfo {year} {2009})}\BibitemShut {NoStop}%
\bibitem [{\citenamefont {Orth}\ \emph {et~al.}(2011)\citenamefont {Orth} \emph
  {et~al.}}]{Orth:2011}%
  \BibitemOpen
  \bibfield  {author} {\bibinfo {author} {\bibfnamefont {J.~D.}\ \bibnamefont
  {Orth}} \emph {et~al.},\ }\bibfield  {title} {\enquote {\bibinfo {title} {{A
  comprehensive genome-scale reconstruction of {\it Escherichia coli}
  metabolism-2011}},}\ }\href@noop {} {\bibfield  {journal} {\bibinfo
  {journal} {Molecular Systems Biology}\ }\textbf {\bibinfo {volume} {7}},\
  \bibinfo {pages} {535} (\bibinfo {year} {2011})}\BibitemShut {NoStop}%
\bibitem [{\citenamefont {Fong}\ \emph {et~al.}(2003)\citenamefont {Fong},
  \citenamefont {Marciniak},\ and\ \citenamefont {Palsson}}]{Fong:2003}%
  \BibitemOpen
  \bibfield  {author} {\bibinfo {author} {\bibfnamefont {S.}~\bibnamefont
  {Fong}}, \bibinfo {author} {\bibfnamefont {J.~Y.}\ \bibnamefont {Marciniak}},
  \ and\ \bibinfo {author} {\bibfnamefont {B.~{\O}.}\ \bibnamefont {Palsson}},\
  }\bibfield  {title} {\enquote {\bibinfo {title} {{Description and
  interpretation of adaptive evolution of {\it Escherichia coli} K-12 MG1655 by
  using a genome-scale in silico metabolic model}},}\ }\href@noop {} {\bibfield
   {journal} {\bibinfo  {journal} {Journal of Bacteriology}\ }\textbf {\bibinfo
  {volume} {185}},\ \bibinfo {pages} {6400--6408} (\bibinfo {year}
  {2003})}\BibitemShut {NoStop}%
\bibitem [{\citenamefont {Edwards}\ \emph {et~al.}(2001)\citenamefont
  {Edwards}, \citenamefont {Ibarra},\ and\ \citenamefont
  {Palsson}}]{Edwards2001}%
  \BibitemOpen
  \bibfield  {author} {\bibinfo {author} {\bibfnamefont {J.~S.}\ \bibnamefont
  {Edwards}}, \bibinfo {author} {\bibfnamefont {R.~U.}\ \bibnamefont {Ibarra}},
  \ and\ \bibinfo {author} {\bibfnamefont {B.~{\O}.}\ \bibnamefont {Palsson}},\
  }\bibfield  {title} {\enquote {\bibinfo {title} {{{\it In silico} predictions
  of {\it Escherichia coli} metabolic capabilities are consistent with
  experimental data}},}\ }\href@noop {} {\bibfield  {journal} {\bibinfo
  {journal} {Nature Biotechnology}\ }\textbf {\bibinfo {volume} {19}},\
  \bibinfo {pages} {125--130} (\bibinfo {year} {2001})}\BibitemShut {NoStop}%
\bibitem [{\citenamefont {Varma}\ \emph {et~al.}(1993)\citenamefont {Varma},
  \citenamefont {Boesch},\ and\ \citenamefont {Palsson}}]{Varma:1993}%
  \BibitemOpen
  \bibfield  {author} {\bibinfo {author} {\bibfnamefont {A.}~\bibnamefont
  {Varma}}, \bibinfo {author} {\bibfnamefont {B.~W.}\ \bibnamefont {Boesch}}, \
  and\ \bibinfo {author} {\bibfnamefont {B.~{\O}.}\ \bibnamefont {Palsson}},\
  }\bibfield  {title} {\enquote {\bibinfo {title} {{Stoichiometric
  interpretation of {\it Escherichia coli} glucose catabolism under various
  oxygenation rates}},}\ }\href@noop {} {\bibfield  {journal} {\bibinfo
  {journal} {Applied and Environmental Microbiology}\ }\textbf {\bibinfo
  {volume} {59}},\ \bibinfo {pages} {2465--2473} (\bibinfo {year}
  {1993})}\BibitemShut {NoStop}%
\bibitem [{\citenamefont {Reiling}\ \emph {et~al.}(1985)\citenamefont
  {Reiling}, \citenamefont {Laurila},\ and\ \citenamefont
  {Fiechter}}]{Reiling:1985}%
  \BibitemOpen
  \bibfield  {author} {\bibinfo {author} {\bibfnamefont {H.~E.}\ \bibnamefont
  {Reiling}}, \bibinfo {author} {\bibfnamefont {H.}~\bibnamefont {Laurila}}, \
  and\ \bibinfo {author} {\bibfnamefont {A.}~\bibnamefont {Fiechter}},\
  }\bibfield  {title} {\enquote {\bibinfo {title} {{Mass culture of {\it
  Escherichia coli}: medium development for low and high density cultivation of
  {\it Escherichia coli} B/r in minimal and complex media}},}\ }\href@noop {}
  {\bibfield  {journal} {\bibinfo  {journal} {Journal of Biotechnology}\
  }\textbf {\bibinfo {volume} {2}},\ \bibinfo {pages} {191--206} (\bibinfo
  {year} {1985})}\BibitemShut {NoStop}%
\bibitem [{\citenamefont {El-Mansi}\ and\ \citenamefont
  {Holms}(1989)}]{El-Mansi:1989}%
  \BibitemOpen
  \bibfield  {author} {\bibinfo {author} {\bibfnamefont {E.~M.}\ \bibnamefont
  {El-Mansi}}\ and\ \bibinfo {author} {\bibfnamefont {W.~H.}\ \bibnamefont
  {Holms}},\ }\bibfield  {title} {\enquote {\bibinfo {title} {{Control of
  carbon flux to acetate excretion during growth of {\it Escherichia coli} in
  batch and continuous cultures}},}\ }\href@noop {} {\bibfield  {journal}
  {\bibinfo  {journal} {Journal of General Microbiology}\ }\textbf {\bibinfo
  {volume} {135}},\ \bibinfo {pages} {2875--2883} (\bibinfo {year}
  {1989})}\BibitemShut {NoStop}%
\bibitem [{\citenamefont {Wolfe}(2005)}]{Wolfe:2005}%
  \BibitemOpen
  \bibfield  {author} {\bibinfo {author} {\bibfnamefont {A.~J.}\ \bibnamefont
  {Wolfe}},\ }\bibfield  {title} {\enquote {\bibinfo {title} {The acetate
  switch},}\ }\href@noop {} {\bibfield  {journal} {\bibinfo  {journal}
  {Microbiology and Molecular Biology Reviews}\ }\textbf {\bibinfo {volume}
  {69}},\ \bibinfo {pages} {12--50} (\bibinfo {year} {2005})}\BibitemShut
  {NoStop}%
\bibitem [{\citenamefont {Varman}\ \emph {et~al.}(2014)\citenamefont {Varman},
  \citenamefont {He}, \citenamefont {You}, \citenamefont {Hollinshead},\ and\
  \citenamefont {Tang}}]{Varman:2014}%
  \BibitemOpen
  \bibfield  {author} {\bibinfo {author} {\bibfnamefont {A.~M.}\ \bibnamefont
  {Varman}}, \bibinfo {author} {\bibfnamefont {L.}~\bibnamefont {He}}, \bibinfo
  {author} {\bibfnamefont {L.}~\bibnamefont {You}}, \bibinfo {author}
  {\bibfnamefont {W.}~\bibnamefont {Hollinshead}}, \ and\ \bibinfo {author}
  {\bibfnamefont {Y.~J.}\ \bibnamefont {Tang}},\ }\bibfield  {title} {\enquote
  {\bibinfo {title} {Elucidation of intrinsic biosynthesis yields using
  13c-based metabolism analysis},}\ }\href@noop {} {\bibfield  {journal}
  {\bibinfo  {journal} {Microbial Cell Factories}\ }\textbf {\bibinfo {volume}
  {13}},\ \bibinfo {pages} {42} (\bibinfo {year} {2014})}\BibitemShut {NoStop}%
\bibitem [{\citenamefont {V\'{a}zquez}\ \emph {et~al.}(2008)\citenamefont
  {V\'{a}zquez}, \citenamefont {Beg}, \citenamefont {deMenezes}, \citenamefont
  {Ernst}, \citenamefont {Bar-Joseph}, \citenamefont {Barab{\'a}si},
  \citenamefont {Boros},\ and\ \citenamefont {Oltvai}}]{Vazquez:2008}%
  \BibitemOpen
  \bibfield  {author} {\bibinfo {author} {\bibfnamefont {A.}~\bibnamefont
  {V\'{a}zquez}}, \bibinfo {author} {\bibfnamefont {Q.~K.}\ \bibnamefont
  {Beg}}, \bibinfo {author} {\bibfnamefont {M.~A.}\ \bibnamefont {deMenezes}},
  \bibinfo {author} {\bibfnamefont {J.}~\bibnamefont {Ernst}}, \bibinfo
  {author} {\bibfnamefont {Z.}~\bibnamefont {Bar-Joseph}}, \bibinfo {author}
  {\bibfnamefont {A.~L.}\ \bibnamefont {Barab{\'a}si}}, \bibinfo {author}
  {\bibfnamefont {L.~G.}\ \bibnamefont {Boros}}, \ and\ \bibinfo {author}
  {\bibfnamefont {Z.~N.}\ \bibnamefont {Oltvai}},\ }\bibfield  {title}
  {\enquote {\bibinfo {title} {{Impact of the solvent capacity constraint on
  {\it E. coli} metabolism}},}\ }\href@noop {} {\bibfield  {journal} {\bibinfo
  {journal} {BMC Systems Biology}\ }\textbf {\bibinfo {volume} {2}},\ \bibinfo
  {pages} {7} (\bibinfo {year} {2008})}\BibitemShut {NoStop}%
\bibitem [{\citenamefont {Varma}\ and\ \citenamefont
  {Palsson}(1994)}]{Varma1994b}%
  \BibitemOpen
  \bibfield  {author} {\bibinfo {author} {\bibfnamefont {A.}~\bibnamefont
  {Varma}}\ and\ \bibinfo {author} {\bibfnamefont {B.~{\O}}\ \bibnamefont
  {Palsson}},\ }\bibfield  {title} {\enquote {\bibinfo {title} {{Stoichiometric
  flux balance models quantitatively predict growth and metabolic by-product
  secretion in wild-type {\it Escherichia coli} W3110}},}\ }\href@noop {}
  {\bibfield  {journal} {\bibinfo  {journal} {Applied and Environmental
  Microbiology}\ }\textbf {\bibinfo {volume} {60}},\ \bibinfo {pages}
  {3724--3731} (\bibinfo {year} {1994})}\BibitemShut {NoStop}%
\bibitem [{\citenamefont {Molenaar}\ \emph {et~al.}(2009)\citenamefont
  {Molenaar}, \citenamefont {van Berlo}, \citenamefont {de~Ridder},\ and\
  \citenamefont {Teusink}}]{Molenaar:2009}%
  \BibitemOpen
  \bibfield  {author} {\bibinfo {author} {\bibfnamefont {D.}~\bibnamefont
  {Molenaar}}, \bibinfo {author} {\bibfnamefont {R.}~\bibnamefont {van Berlo}},
  \bibinfo {author} {\bibfnamefont {D.}~\bibnamefont {de~Ridder}}, \ and\
  \bibinfo {author} {\bibfnamefont {B.}~\bibnamefont {Teusink}},\ }\bibfield
  {title} {\enquote {\bibinfo {title} {Shifts in growth strategies reflect
  tradeoffs in cellular economics},}\ }\href@noop {} {\bibfield  {journal}
  {\bibinfo  {journal} {Molecular Systems Biology}\ }\textbf {\bibinfo {volume}
  {5}} (\bibinfo {year} {2009})}\BibitemShut {NoStop}%
\bibitem [{\citenamefont {Wortel}\ \emph {et~al.}(2014)\citenamefont {Wortel},
  \citenamefont {Peters}, \citenamefont {Hulshof}, \citenamefont {Teusink},\
  and\ \citenamefont {Bruggeman}}]{Wortel:2014}%
  \BibitemOpen
  \bibfield  {author} {\bibinfo {author} {\bibfnamefont {M.~T.}\ \bibnamefont
  {Wortel}}, \bibinfo {author} {\bibfnamefont {H.}~\bibnamefont {Peters}},
  \bibinfo {author} {\bibfnamefont {J.}~\bibnamefont {Hulshof}}, \bibinfo
  {author} {\bibfnamefont {B.}~\bibnamefont {Teusink}}, \ and\ \bibinfo
  {author} {\bibfnamefont {F.~J.}\ \bibnamefont {Bruggeman}},\ }\bibfield
  {title} {\enquote {\bibinfo {title} {Metabolic states with maximal specific
  rate carry flux through an elementary flux mode},}\ }\href@noop {} {\bibfield
   {journal} {\bibinfo  {journal} {FEBS Journal}\ }\textbf {\bibinfo {volume}
  {281}},\ \bibinfo {pages} {1547--1555} (\bibinfo {year} {2014})}\BibitemShut
  {NoStop}%
\bibitem [{\citenamefont {Chen}\ \emph {et~al.}(2011)\citenamefont {Chen},
  \citenamefont {Alonso}, \citenamefont {Allen}, \citenamefont {Reed},\ and\
  \citenamefont {Shachar-Hill}}]{Chen:2011}%
  \BibitemOpen
  \bibfield  {author} {\bibinfo {author} {\bibfnamefont {X.}~\bibnamefont
  {Chen}}, \bibinfo {author} {\bibfnamefont {A.~P.}\ \bibnamefont {Alonso}},
  \bibinfo {author} {\bibfnamefont {D.~K.}\ \bibnamefont {Allen}}, \bibinfo
  {author} {\bibfnamefont {J.~L.}\ \bibnamefont {Reed}}, \ and\ \bibinfo
  {author} {\bibfnamefont {Y.}~\bibnamefont {Shachar-Hill}},\ }\bibfield
  {title} {\enquote {\bibinfo {title} {Synergy between 13c-metabolic flux
  analysis and flux balance analysis for understanding metabolic adaptation to
  anaerobiosis in {\em e. coli}},}\ }\href@noop {} {\bibfield  {journal}
  {\bibinfo  {journal} {Metabolic Engineering}\ }\textbf {\bibinfo {volume}
  {13}},\ \bibinfo {pages} {38--48} (\bibinfo {year} {2011})}\BibitemShut
  {NoStop}%
\bibitem [{\citenamefont {Losen}\ \emph {et~al.}(2004)\citenamefont {Losen},
  \citenamefont {Frolich}, \citenamefont {Pohl},\ and\ \citenamefont
  {Buchs}}]{Losen:2004}%
  \BibitemOpen
  \bibfield  {author} {\bibinfo {author} {\bibfnamefont {M.}~\bibnamefont
  {Losen}}, \bibinfo {author} {\bibfnamefont {B.}~\bibnamefont {Frolich}},
  \bibinfo {author} {\bibfnamefont {M.}~\bibnamefont {Pohl}}, \ and\ \bibinfo
  {author} {\bibfnamefont {J.}~\bibnamefont {Buchs}},\ }\bibfield  {title}
  {\enquote {\bibinfo {title} {{Effect of oxygen limitation and medium
  composition on {\it Escherichia coli} fermentation in shake-flask
  cultures}},}\ }\href@noop {} {\bibfield  {journal} {\bibinfo  {journal}
  {Biotechnology Progress}\ }\textbf {\bibinfo {volume} {20}},\ \bibinfo
  {pages} {1062--1068} (\bibinfo {year} {2004})}\BibitemShut {NoStop}%
\bibitem [{\citenamefont {Chubukov}\ \emph {et~al.}(2014)\citenamefont
  {Chubukov}, \citenamefont {Gerosa}, \citenamefont {Kochanowski},\ and\
  \citenamefont {Sauer}}]{Chubukov:2014}%
  \BibitemOpen
  \bibfield  {author} {\bibinfo {author} {\bibfnamefont {Victor}\ \bibnamefont
  {Chubukov}}, \bibinfo {author} {\bibfnamefont {Luca}\ \bibnamefont {Gerosa}},
  \bibinfo {author} {\bibfnamefont {Karl}\ \bibnamefont {Kochanowski}}, \ and\
  \bibinfo {author} {\bibfnamefont {Uwe}\ \bibnamefont {Sauer}},\ }\bibfield
  {title} {\enquote {\bibinfo {title} {Coordination of microbial metabolism},}\
  }\href@noop {} {\bibfield  {journal} {\bibinfo  {journal} {Nature Reviews}\
  }\textbf {\bibinfo {volume} {12}},\ \bibinfo {pages} {327--340} (\bibinfo
  {year} {2014})}\BibitemShut {NoStop}%
\bibitem [{\citenamefont {Orencio-Trejo}\ \emph {et~al.}(2010)\citenamefont
  {Orencio-Trejo}, \citenamefont {J.~Utrilla}, \citenamefont
  {Huerta-Beristain}, \citenamefont {Gosset},\ and\ \citenamefont
  {Martinez}}]{Orencio-Trejo:2010}%
  \BibitemOpen
  \bibfield  {author} {\bibinfo {author} {\bibfnamefont {M.}~\bibnamefont
  {Orencio-Trejo}}, \bibinfo {author} {\bibfnamefont {M.T.
  Fern\'andez-Sandoval}\ \bibnamefont {J.~Utrilla}}, \bibinfo {author}
  {\bibfnamefont {G.}~\bibnamefont {Huerta-Beristain}}, \bibinfo {author}
  {\bibfnamefont {G.}~\bibnamefont {Gosset}}, \ and\ \bibinfo {author}
  {\bibfnamefont {A.}~\bibnamefont {Martinez}},\ }\bibfield  {title} {\enquote
  {\bibinfo {title} {{Engineering the {\it Escherichia coli} fermentative
  metabolism}},}\ }\href@noop {} {\bibfield  {journal} {\bibinfo  {journal}
  {Advances in Biochemical Engineering/Biotechnology}\ }\textbf {\bibinfo
  {volume} {121}},\ \bibinfo {pages} {71--107} (\bibinfo {year}
  {2010})}\BibitemShut {NoStop}%
\bibitem [{\citenamefont {Price}\ and\ \citenamefont
  {Shmulevich}(2007)}]{Price:2007}%
  \BibitemOpen
  \bibfield  {author} {\bibinfo {author} {\bibfnamefont {N.~D.}\ \bibnamefont
  {Price}}\ and\ \bibinfo {author} {\bibfnamefont {I.}~\bibnamefont
  {Shmulevich}},\ }\bibfield  {title} {\enquote {\bibinfo {title} {Biochemical
  and statistical network models for systems biology},}\ }\href@noop {}
  {\bibfield  {journal} {\bibinfo  {journal} {Current Opinion in
  Biotechnology}\ }\textbf {\bibinfo {volume} {18}},\ \bibinfo {pages}
  {365--370} (\bibinfo {year} {2007})}\BibitemShut {NoStop}%
\bibitem [{\citenamefont {Folger}\ \emph {et~al.}(2011)\citenamefont {Folger},
  \citenamefont {Jerby}, \citenamefont {Frezza}, \citenamefont {Gottlieb},
  \citenamefont {Ruppin},\ and\ \citenamefont {Shlomi}}]{Folger:2011}%
  \BibitemOpen
  \bibfield  {author} {\bibinfo {author} {\bibfnamefont {O.}~\bibnamefont
  {Folger}}, \bibinfo {author} {\bibfnamefont {L.}~\bibnamefont {Jerby}},
  \bibinfo {author} {\bibfnamefont {C.}~\bibnamefont {Frezza}}, \bibinfo
  {author} {\bibfnamefont {E.}~\bibnamefont {Gottlieb}}, \bibinfo {author}
  {\bibfnamefont {E.}~\bibnamefont {Ruppin}}, \ and\ \bibinfo {author}
  {\bibfnamefont {T.}~\bibnamefont {Shlomi}},\ }\bibfield  {title} {\enquote
  {\bibinfo {title} {Predicting selective drug targets in cancer through
  metabolic networks},}\ }\href@noop {} {\bibfield  {journal} {\bibinfo
  {journal} {Molecular Systems Biology}\ }\textbf {\bibinfo {volume} {7}},\
  \bibinfo {pages} {501} (\bibinfo {year} {2011})}\BibitemShut {NoStop}%
\end{thebibliography}


\newpage
\section{Supplementary Information}

\renewcommand{\figurename}{Figure S}
\renewcommand{\tablename}{Table S}

\section{Figure S1}
\begin{figure}[ht!]
 \addtocounter{figure}{-3}
 \centering
 \includegraphics[width=\textwidth]{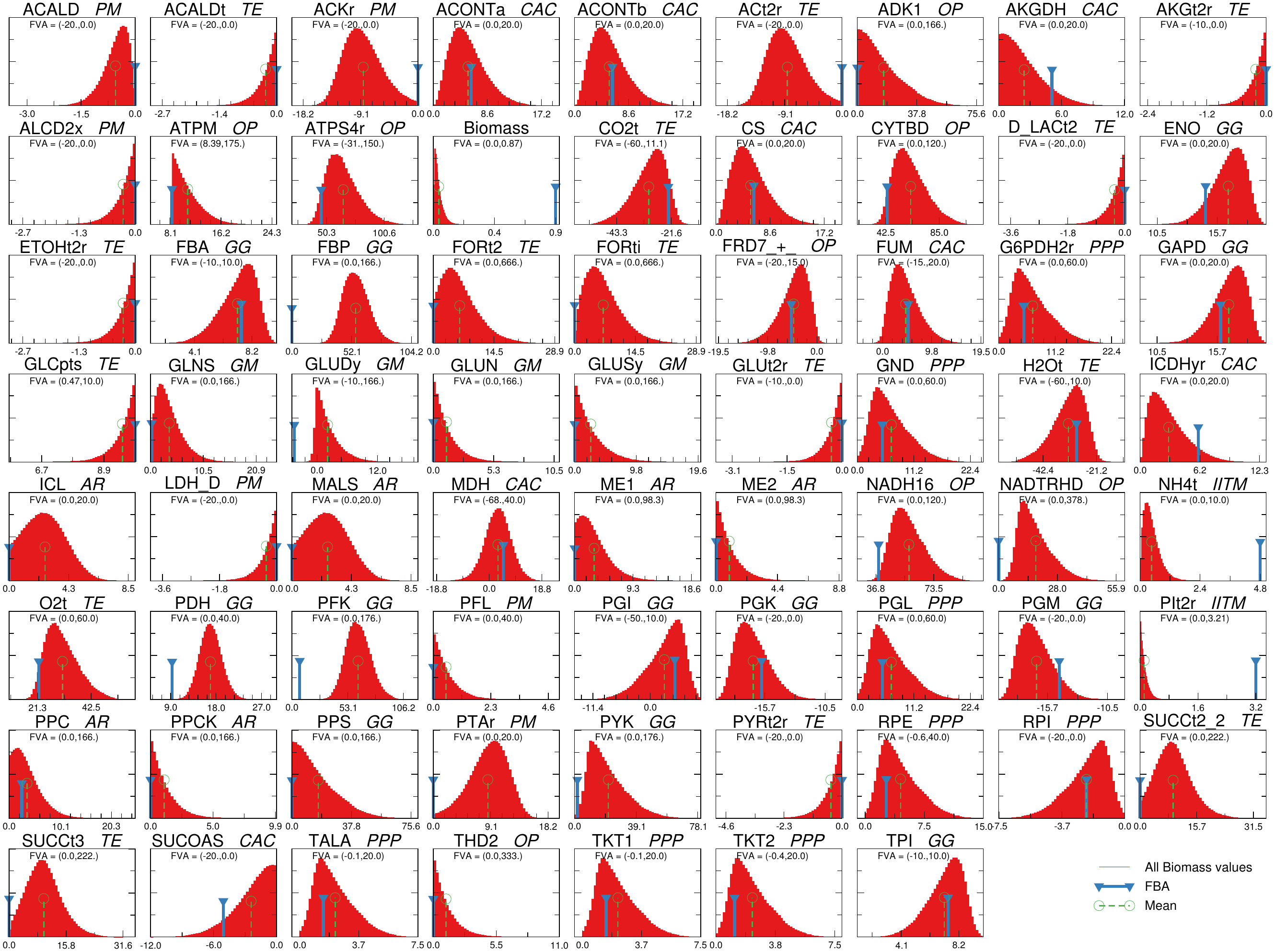}
 \caption{Probability density functions of metabolic fluxes values for all reactions in core {\it E. coli} under glucose minimal conditions. Each graph shows the reaction label, the flux variability range (values inside parentheses), and each associated pathway (acronyms in italics). Notice that the range plotted in the axes does not coincide with the flux variability range since in the axes we chose an optimal x range for each reaction to distinguish the shape of each profile. In addition, in each profile we also show the position of the FBA point (blue marker) and the position of the Mean (green marker).}
 \label{fig:figs1}
\end{figure}

\newpage
\section{Figure S2}

\begin{figure}[ht!]
 \centering
 \includegraphics[width=\textwidth]{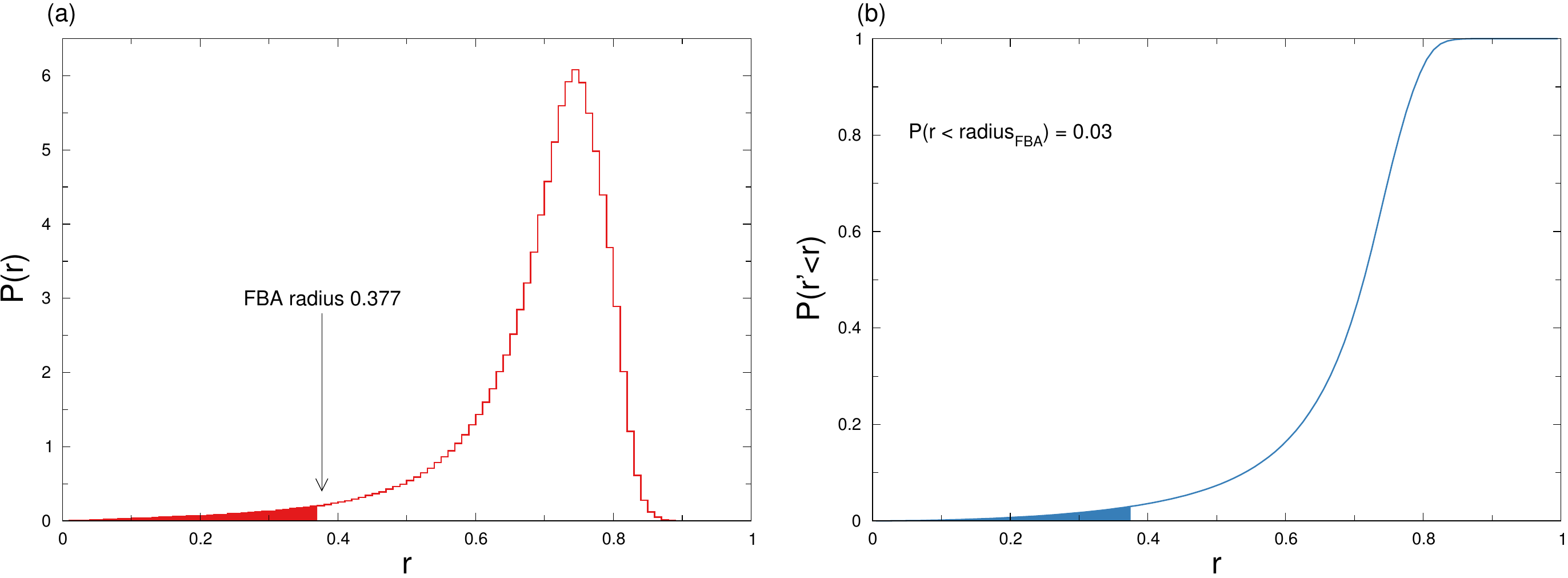}
 \caption{\textbf{(a)} Probability distribution function of the radii of all solutions before applying the negative logarithmic transformation. The red area denotes the probability of having a smaller radius than the FBA solution. This fraction of area is the 3\% of the total area, which means that the 97\% of the solutions have a larger radius than the FBA solution. \textbf{(b)} Cumulative probability distribution function of the radius. The blue region denotes the range of solutions with a radius smaller than FBA. The probability of having a radius smaller than FBA is the y-value of the curve at the rightmost side of the region.}
 \label{fig:figs2}
\end{figure}

\newpage
\section{Figure S3}

\begin{figure}[ht!]
 \centering
 \includegraphics[width=0.6\textwidth]{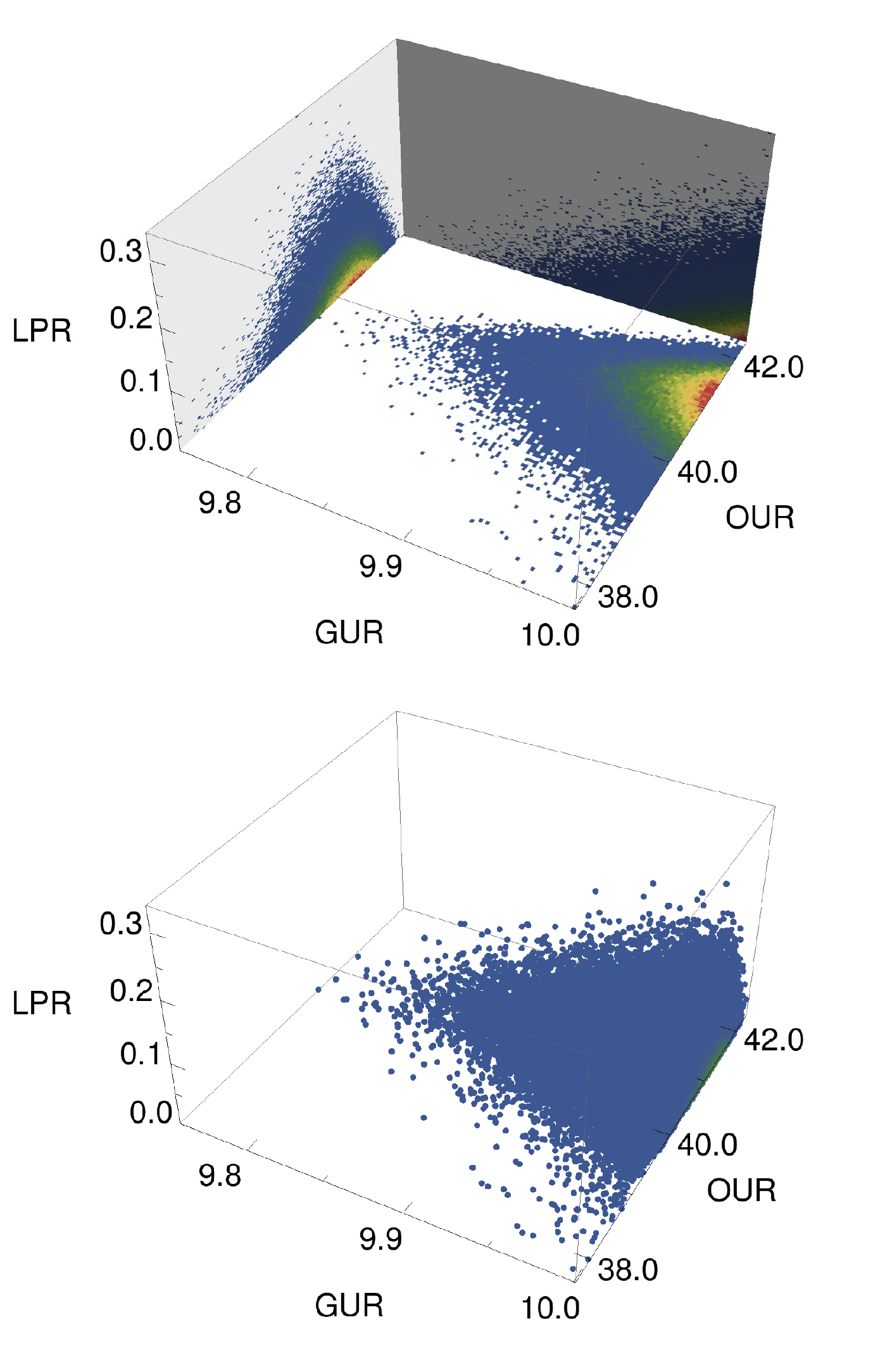}
 \caption{Lactate high growth phenotype of core {\it E. coli} on glucose. Up: Density projection of the 3-dimensional scatterplot on each of the possible 2D planes lactate-glucose, lactate-oxygen, and glucose-oxygen. Bottom: 3-dimensional scatterplot of lactate production rate vs glucose and oxygen uptake rates.}
 \label{fig:figs3}
\end{figure}

\newpage
\section{Figure S4}
It was proven~[1] that, by iterating steps (1-4) of the ``Hit-And-Run algotihm to sample the space of feasible metabolic flux solutions" section in the main text, the samples obtained are asymptotically unbiased, in the sense that the whole FFP space is explored with the same likelihood, in the limit of very large samples. In practice, one must always work with a finite sample, and hence we have taken some additional measures to ensure that our samples were truly representative of the whole FFP space. In particular:

\begin{enumerate}
\item Only one every $10^3$ points generated by HR was included in the final sample. This effectively decreases the ``mixing time'' of the algorithm, since the correlation among the points that are actually retained decays fast.
\item Different initial conditions were used. Results showed no dependence on the initial condition, as expected for large samples. Even so, the first 30\% of points was discarded, in order to rule out any subtler effect of the initial condition on the final results.
\item Results were recalculated using subsamples of size 10\% of the original sample. We did not find any qualitative difference between the two sets.
\end{enumerate}

Because the HR algorithm is very efficient itself and due to the dimensionality reduction that our implementation adds (see~[2] for details), we were able to generate very large samples in reasonable time. For each model, we initially created samples of size $ 1.3 \cdot10^9$, giving rise to a final set of $10^6$ feasible solutions, uniformly distributed along the whole FFP space.

\begin{figure}[ht!]
 \centering
 \includegraphics[width=0.6\textwidth]{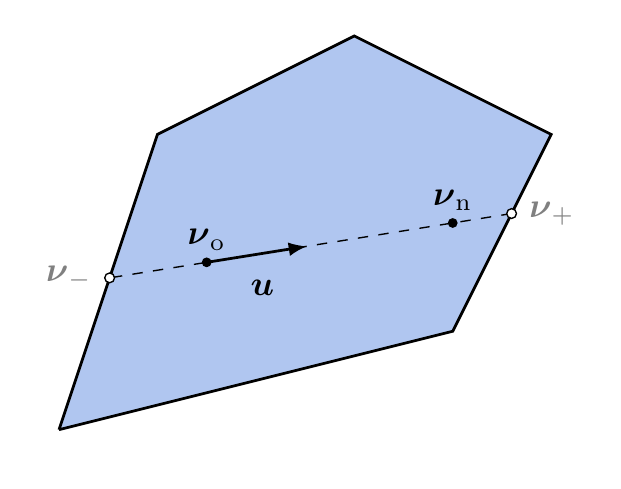}
 \includegraphics[width=0.6\textwidth]{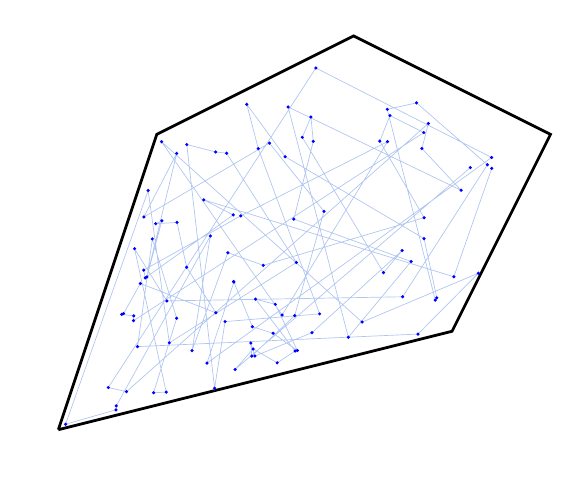}
 \caption{Illustrative representation of the HR fundamental step, which generates a new feasible state $\mathbf{\upsilon}_n$ from a given one $\mathbf{\upsilon}_o$.}
 \label{fig:figs4}
\end{figure}

\clearpage
\section{Figure S5}

\begin{figure}[ht!]
 \centering
 \includegraphics[width=\textwidth]{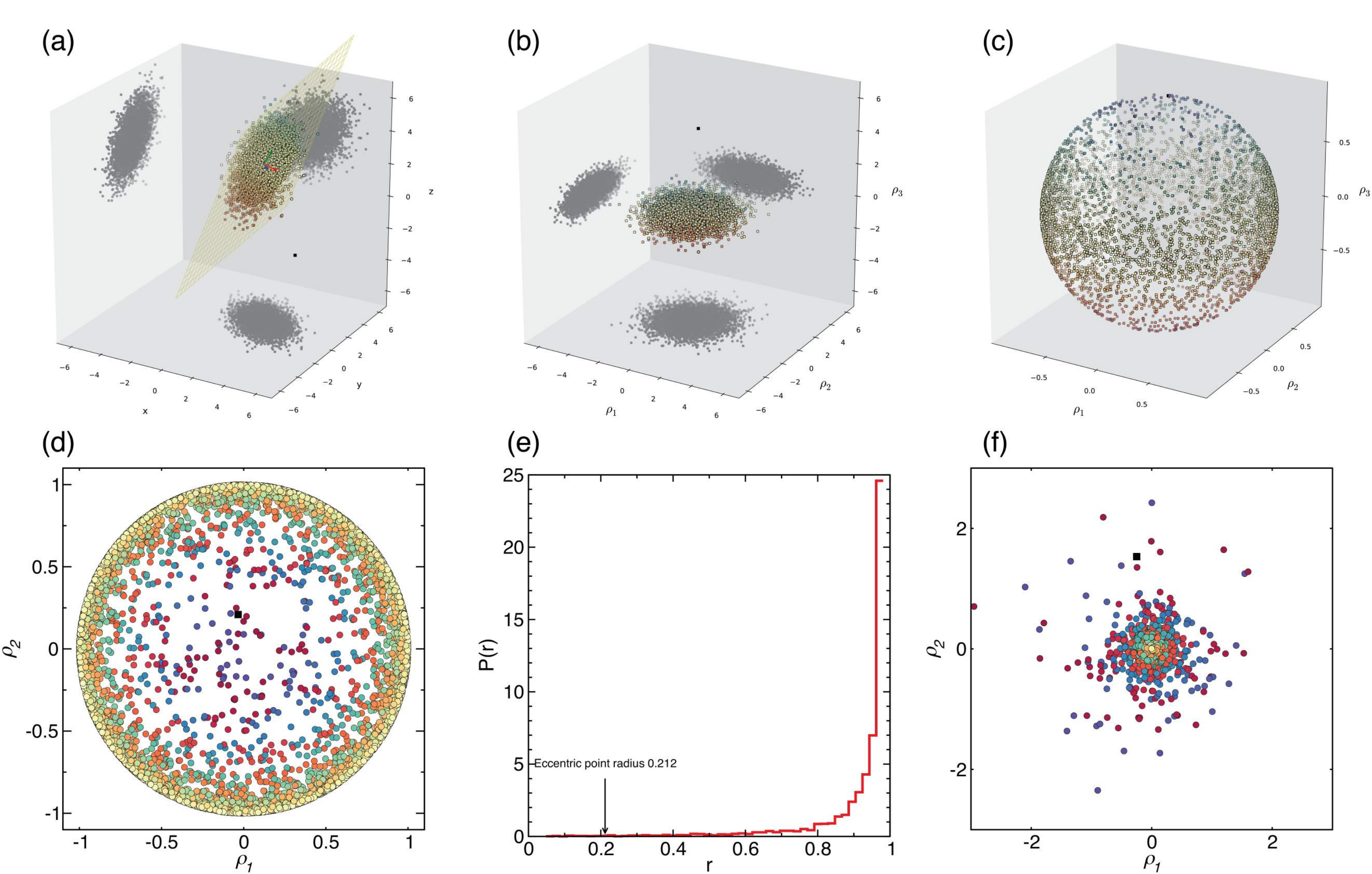}
 \caption{An example to describe our PCA analysis.  {\bf (a)} FFP sampling produces a cloud of points in a multidimensional space that, when projected along the $(x,z)$, $(y,z)$ and $(x,y)$ planes, is seen to span a wide range of values.
Finding the eigenvectors of the correlation matrix, one can see that such points are actually clustered around a plane (plotted as a yellow grid). %
By diagonalizing the ($3\times 3$) correlation matrix, we find the three vectors (plotted in blue, red and green, respectively) identifying the direction in space where the points show most variation, in a decreasing manner. %
We also plot as reference eccentric point a black square. 
{\bf (b)} By projecting the sampled FFP along vectors $\rho_1$, $\rho_2$,  and $\rho_3$, all points are squeezed in a thin region close to the ($\rho_1$, $\rho_2$) plane. %
This shows that the greatest variability of the sampled points actually occurs in the $\rho_1$, $\rho_2$ directions. In this representation, the eccentric black square point is seen to lie far from the plane with a large $\rho_3$ coordinate. %
({\it Follows to next page})}
 \label{fig:figs5}
\end{figure}
\begin{figure}[ht!]
 \addtocounter{figure}{-1}
% \ContinuedFloat 
 \centering
 \caption{({\it Follows from previous page})
{\bf (c)} Normalizing the projection in (b) over the modulus of the vector identifying the point coordinate allows to quantify the closeness to the ($\rho_1$, $\rho_2$) plane. %
In such way all points are projected over the unit radius sphere, with the majority of points scattered near the equator ({\it i.e.} the ($\rho_1$, $\rho_2$) plane). %
Therefore, in this representation, eccentric points like the black square are close to the pole. %
{\bf (d)} Points on the unit sphere may in turn be projected on the ($\rho_1$, $\rho_2$) plane only. In this way all points are constrained within the unitary radius circle, with points close to the equator in plot (c) now close to the circle and the ones close to the pole in (c) near the origin. %
In this representation, typical points ({\it i.e.} those originally closer to the yellow plane in (a)) have larger radius (close to one, but smaller than that) and eccentric points have a smaller radius, like the black square. %
{\bf (e)} Plotting the distribution of the points radius, as in Fig. S3, one sees that $P(r)$ has indeed a peak in one, with very low probability of finding a point with a radius close to zero. Similarly to Fig. S3 we indicate the radius of the eccentric point, highlighting how low $r$, eccentric points are indeed unlikely. %
{\bf (f)} Similarly to Fig. 1f in the main text, we re-project the points on the ($\rho_1$, $\rho_2$) plane, but with a negative log radius. Here all points plotted in panel (d) appear with the same angular coordinate they have in (d) but with a radius $r' = -\log(r)$. In this way, typical points that in (d) have almost unitary radius now coalesce towards the origin and atypical points, that in (d) lie close to 0, are now pushed away from the origin, like the black square. A similar pattern is observed in Fig. 1f in the main text, where the majority of points converge towards the origin and FBA is seen to be a rather eccentric outlier.}
 \label{fig:figs5b}
\end{figure}

\clearpage
\newpage

\section{Figure S6}

\begin{figure}[ht!]
\centering
\includegraphics[width=\textwidth]{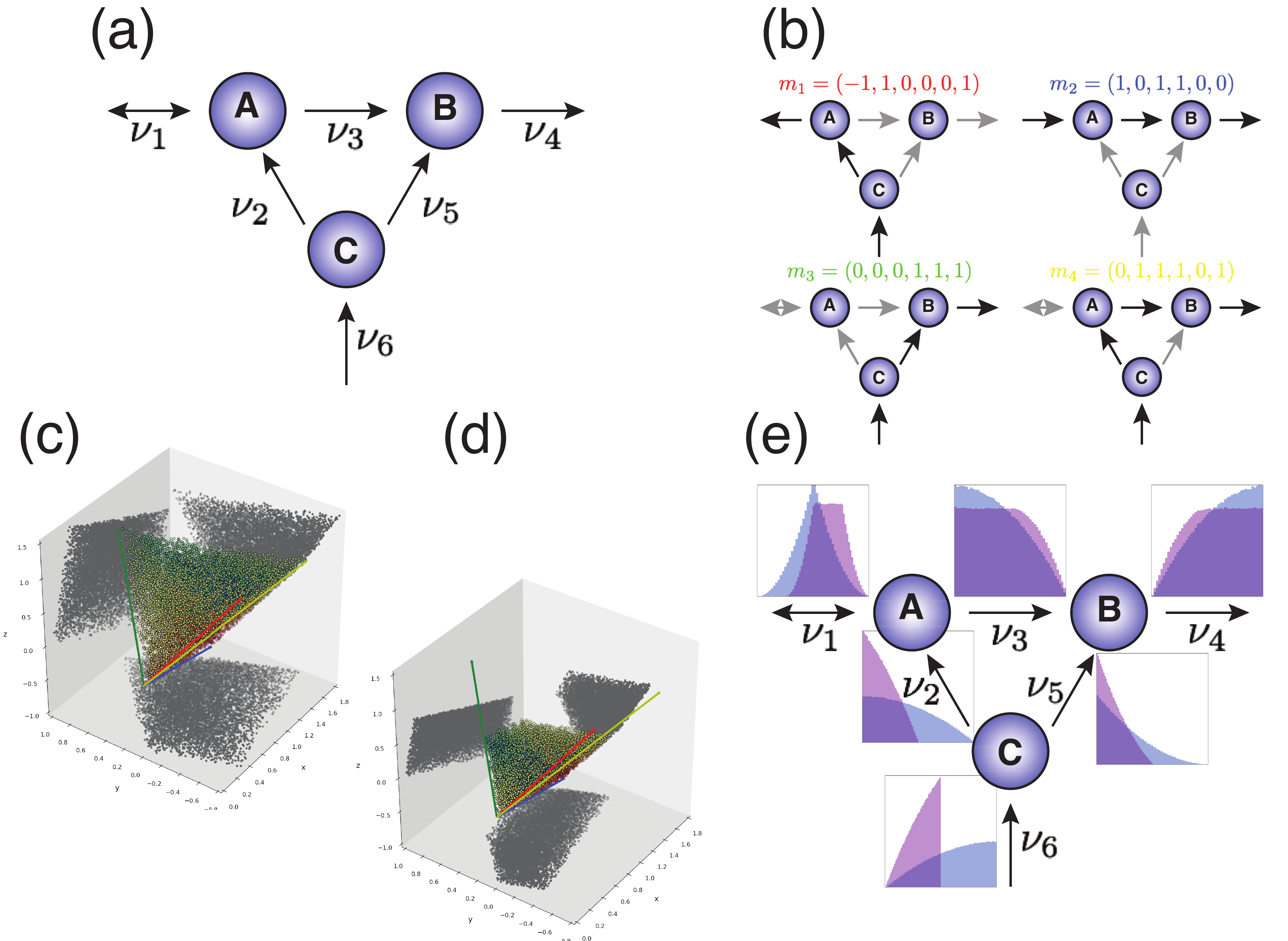}
\caption{Elementary flux modes vs FFP sampling. {\bf (a)} A toy example of a metabolic network adapted from~[3], comprising 3 metabolites ($A$, $B$, $C$) and 6 reactions ($\nu_i$, $i=1,\ldots, 6$). {\bf (b)} The four elementary flux modes (EFMs) of the network $m_1, m_2, m_3, m_4$ are simple paths connecting metabolic inputs to outputs. Their vectorial representation is also shown on top of each graph. Note that these modes depend on the stoichiometry of the network and do not capture, per se, reactions being capped somehow. {\bf (c)} Visualization of the FFP space and EFMs. Since the system features $N=6$ reactions and $M=3$ metabolites, the FFP space has dimension $N-M = 3$ in this case and can be thus visualized in a 3D basis spanning the kernel of the stoichiometric matrix. 
({\it Follows to next page})}
\label{fig:figs_efm}
\end{figure}
\begin{figure}[t!]
 \addtocounter{figure}{-1}
 %\ContinuedFloat 
 \centering
 \caption{({\it Follows from previous page})
All feasible fluxes ({\it i.e.} the FFP space) occupy a `flux cone' whose edges are spanned by the elementary flux modes [the colors of each mode are the same as in (b)], which in this basis read $\hat{m}_1 = (0.3,  -0.81,  1.5)$, 
$\hat{m}_2 = (1.5,   0.3,  -0.81)$, 
$\hat{m}_3 =  (1. ,   1.,    1.  )$, and 
$\hat{m}_4 = ( 1.81, -0.5,   0.7 )$. 
All feasible fluxes are therefore a linear combination of vectors $\hat{m}_\kappa, \kappa=1,\ldots 4$, but these combinations are not unique and, in addition, without knowledge of the whole FFP space it is impossible to tell the likelihood of a particular combination. In this example we chose all reaction fluxes to be in the range $[0,1]$ except for $\nu_1 \in [-1,1]$ being a reversible reaction. {\bf (d)} The effect of capping a reaction on the FFP space. In this case, we capped $\nu_6 \in [0,0.5]$: this changes dramatically the shape of the FFP space as can be appreciated also by observing the shadow of the sampled points on the boundaries of the graph. Although the space is profoundly changed, the EFMs are still the same as in case (c). {\bf (e)} Sampling the FFP space allows to directly assess the likelihood of a given flux value. Expressing back the flux cone in the full 6--dimensional space of reactions and projecting this volume along each of the 6 axes allows to assess the probability of a given flux value, for each reaction. We show here the probability distributions for all reactions, for the case $\nu_6 \in [0,1]$ (light blue) and for $\nu_6 \in [0,0.5]$ (light purple) near each reaction arrow (graph scales are not all the same). The analysis of the whole FFP space allows to see that by varying the upper bound of $\nu_6$, the probability of observing determinate flux values changes drastically, something that EFM analysis does not directly allow. Viceversa, knowing which reactions are more likely to be active allows to easily detect the impact of diverse EFMs on typical metabolic behavior.
}
\label{fig:figs_efmb}
\end{figure}

\clearpage

\newpage

\section{Table S1}

\begin{table}[ht!]
\centering
\begin{tabular}{c | p{4.5cm} | c | c | c}
Reaction & Official name & $CV(f(v))$ & $\frac{d_{PL}}{\nu_U-\nu_L}$ & $\frac{d_{FM}}{\nu_U-\nu_L}$\\
\hline
ACALD	&	acetaldehyde dehydrogenase (acetylating)	&	0.63	&	0.98	&	27	\\
ACALDt	&	acetaldehyde reversible transport	&	0.92	&	1.00	&	13	\\
ACKr	&	acetate kinase	&	0.34	&	0.53	&	0.43	\\
ACONTa	&	aconitase (half-reaction A, Citrate hydro-lyase)	&	0.49	&	0.21	&	25	\\
ACONTb	&	aconitase (half-reaction B, Isocitrate hydro-lyase)	&	0.49	&	0.21	&	25	\\
ACt2r	&	acetate reversible transport via proton symport	&	0.34	&	0.53	&	0.43	\\
ADK1	&	adenylate kinase	&	0.77	&	0.0025	&	0.10	\\
AKGDH	&	2-Oxogluterate dehydrogenase	&	0.76	&	27	&	0.13	\\
AKGt2r	&	2-oxoglutarate reversible transport via symport	&	0.94	&	1.00	&	21	\\
ALCD2x	&	alcohol dehydrogenase (ethanol)	&	0.93	&	1.00	&	14	\\
ATPM	&	ATP maintenance requirement	&	0.19	&	0.00050	&	15	\\
ATPS4r	&	ATP synthase (four protons for one ATP)	&	0.21	&	0.49	&	0.11	\\
Biomass\_Ecoli\_core\_w\_GAM	&	Biomass Objective Function with GAM	&	0.72	&	17	&	0.95	\\
\end{tabular}
\label{fig:t2a1}
\end{table}

\newpage

\begin{table}[ht!]
\centering
\begin{tabular}{c | p{4.5cm} | c | c | c}
Reaction & Official name & $CV(f(v))$ & $\frac{d_{PL}}{\nu_U-\nu_L}$ & $\frac{d_{FM}}{\nu_U-\nu_L}$\\
\hline
CO2t	&	CO2 transporter via diffusion	&	0.21	&	0.46	&	0.11	\\
CS	&	citrate synthase	&	0.49	&	0.21	&	25	\\
CYTBD	&	cytochrome oxidase bd (ubiquinol-8: 2 protons)	&	0.20	&	0.46	&	0.16	\\
D\_LACt2	&	D-lactate transport via proton symport	&	0.94	&	1.00	&	16	\\
ENO	&	enolase	&	87	&	0.87	&	0.10	\\
ETOHt2r	&	ethanol reversible transport via proton symport	&	0.93	&	1.00	&	14	\\
FBA	&	fructose-bisphosphate aldolase	&	0.18	&	0.90	&	13	\\
FBP	&	fructose-bisphosphatase	&	0.20	&	0.33	&	0.33	\\
FORt2	&	formate transport via proton symport (uptake only)	&	0.68	&	0.0055	&	0.0095	\\
FORti	&	formate transport via diffusion	&	0.62	&	0.0065	&	10	\\
FRD7+SUCDi	&	fumarate reductase + succinate dehydrogenase (irreversible)	&	0.60	&	0.47	&	14	\\
FUM	&	fumarase	&	0.60	&	0.53	&	14	\\
G6PDH2r	&	glucose 6-phosphate dehydrogenase	&	0.55	&	70	&	30	\\
GAPD	&	glyceraldehyde-3-phosphate dehydrogenase	&	87	&	0.86	&	36	\\
\end{tabular}
\label{fig:t2a2}
\end{table}

\newpage

\begin{table}[ht!]
\centering
\begin{tabular}{c | p{4.5cm} | c | c | c}
Reaction & Official name & $CV(f(v))$ & $\frac{d_{PL}}{\nu_U-\nu_L}$ & $\frac{d_{FM}}{\nu_U-\nu_L}$\\
\hline
GLCpts	&	D-glucose transport via PEP:Pyr PTS	&	45	&	1.00	&	48	\\
GLNS	&	glutamine synthetase	&	0.67	&	12	&	21	\\
GLUDy	&	glutamate dehydrogenase (NADP)	&	1.13	&	57	&	38	\\
GLUN	&	glutaminase	&	0.94	&	0.00050	&	0.0070	\\
GLUSy	&	glutamate synthase (NADPH)	&	0.91	&	0.00050	&	15	\\
GLUt2r	&	L-glutamate transport via proton symport, reversible (periplasm)	&	0.94	&	1.00	&	29	\\
GND	&	phosphogluconate dehydrogenase	&	0.55	&	70	&	30	\\
H2Ot	&	H2O transport via diffusion	&	0.20	&	0.44	&	46	\\
ICDHyr	&	isocitrate dehydrogenase (NADP)	&	0.62	&	77	&	0.15	\\
ICL	&	Isocitrate lyase	&	0.55	&	0.12	&	0.13	\\
LDH\_D	&	D-lactate dehydrogenase	&	0.94	&	1.00	&	16	\\
MALS	&	malate synthase	&	0.55	&	0.12	&	0.13	\\
MDH	&	malate dehydrogenase	&	1.46	&	0.66	&	18	\\
ME1	&	malic enzyme (NAD)	&	0.76	&	14	&	31	\\
ME2	&	malic enzyme (NADP)	&	0.91	&	0.00050	&	10	\\
\end{tabular}
\end{table}

\newpage

\begin{table}[ht!]
\centering
\begin{tabular}{c | p{4.5cm} | c | c | c}
Reaction & Official name & $CV(f(v))$ & $\frac{d_{PL}}{\nu_U-\nu_L}$ & $\frac{d_{FM}}{\nu_U-\nu_L}$\\
\hline
NADH16	&	NADH dehydrogenase (ubiquinone-8 \& 3 protons)	&	0.18	&	0.45	&	0.16	\\
NADTRHD	&	NAD transhydrogenase	&	0.44	&	31	&	47	\\
NH4t	&	ammonia reversible transport	&	0.67	&	26	&	0.43	\\
O2t	&	o2 transport via diffusion	&	0.20	&	0.46	&	0.16	\\
PDH	&	pyruvate dehydrogenase	&	0.13	&	0.42	&	0.19	\\
PFK	&	phosphofructokinase	&	0.19	&	0.35	&	0.31	\\
PFL	&	pyruvate formate lyase	&	0.91	&	0.00050	&	13	\\
PGI	&	glucose-6-phosphate isomerase	&	1.33	&	0.92	&	35	\\
PGK	&	phosphoglycerate kinase	&	87	&	0.14	&	36	\\
PGL	&	6-phosphogluconolactonase	&	0.55	&	70	&	30	\\
PGM	&	phosphoglycerate mutase	&	87	&	0.13	&	0.10	\\
PIt2r	&	phosphate reversible transport via proton symport	&	0.91	&	0.00050	&	0.96	\\
PPC	&	phosphoenolpyruvate carboxylase	&	0.75	&	10	&	0.0063	\\
PPCK	&	phosphoenolpyruvate carboxykinase	&	0.91	&	0.00050	&	0.0065	\\
PPS	&	phosphoenolpyruvate synthase	&	0.77	&	0.0025	&	0.10	\\
PTAr	&	phosphotransacetylase	&	0.34	&	0.47	&	0.43	\\
\end{tabular}
\label{fig:t2b2}
\end{table}

\newpage

\begin{table}[ht!]
\centering
\begin{tabular}{c | p{4.5cm} | c | c | c}
Reaction & Official name & $CV(f(v))$ & $\frac{d_{PL}}{\nu_U-\nu_L}$ & $\frac{d_{FM}}{\nu_U-\nu_L}$\\
\hline
PYK	&	pyruvate kinase	&	0.56	&	60	&	0.11	\\
PYRt2r	&	pyruvate reversible transport via proton symport	&	0.94	&	1.00	&	22	\\
RPE	&	ribulose 5-phosphate 3-epimerase	&	0.55	&	84	&	45	\\
RPI	&	ribose-5-phosphate isomerase	&	0.54	&	0.93	&	0.00021	\\
SUCCt2\_2	&	succinate transport via proton symport (2 H)	&	0.53	&	39	&	41	\\
SUCCt3	&	succinate transport out via proton antiport	&	0.51	&	42	&	43	\\
SUCOAS	&	succinyl-CoA synthetase (ADP-forming)	&	0.76	&	0.97	&	0.13	\\
TALA	&	transaldolase	&	0.55	&	76	&	37	\\
THD2	&	NAD(P) transhydrogenase	&	0.92	&	0.00050	&	0.0034	\\
TKT1	&	transketolase	&	0.55	&	76	&	37	\\
TKT2	&	transketolase	&	0.53	&	92	&	55	\\
TPI	&	triose-phosphate isomerase	&	0.18	&	0.90	&	13	\\
\end{tabular}
\caption{Table showing the values of $CV(f(\nu))$, $d_{PL}/(\nu_U-\nu_L)$ and $d_{FM}/(\nu_U-\nu_L)$ for each reaction.}
\label{fig:t2c1}
\end{table}

\newpage
\section{Table S2}

\begin{table}[ht!]
 \centering
 \begin{tabular}{c c}
 Acronym & Pathway\\
 \hline
 AR & Anaplerotic reactions\\
 B & Biomass\\
 CAC & Citric Acid Cycle\\
 GM & Glutamate Metabolism\\
 GG & Glycolysis/Gluconeogenesis\\
 IITM & Inorganic Ion Transport and Metabolism\\
 OP & Oxidative Phosphorylation\\
 PPP & Pentose Phosphate Pathway\\
 PM & Pyruvate Metabolism\\
 TE & Transport, Extracellular\\
 \end{tabular}
 \caption{Table showing the acronyms of the pathways used in the main text.}
 \label{fig:t1}
\end{table}

\subsection{Supplementary Information References}

[1] Lov{\'a}sz, L. (1999) {\it Mathematical Programming} {\bf 86}, 443--461.

[2] Massucci, F.~A, Font-Clos, F, Martino, A.~D,  \& Castillo, I.~P. (2013) {\it Metabolites} {\bf 3}, 838--852.

[3] Papin, J.~A., Stelling, J.,  Price, N.~D., Klamt, S., Schuster S. \& Palsson, B.~O. (2004) {\it Trends in Biotechnology} {\bf 8}, 400--405.

\end{document}